\documentclass[11pt,a4paper]{article}
\usepackage{jheppub}
\usepackage{amsmath}
\usepackage{multicol,bbm}
\usepackage{enumerate}
\usepackage{bibentry}
\usepackage[toc,page]{appendix}
\graphicspath{{./Figs/}}

\newcommand{\beqs}{\begin{equation*}}
\def\beq{\begin{equation}}

\newcommand{\eeqs}{\end{equation*}}
\def\eeq{\end{equation}}

\def\beq{\begin{equation}}
\def\eeq{\end{equation}}
\def\be{\begin{equation}}
\def\be{\begin{equation}}
\def\ee{\end{equation}}
\def\ba{\begin{eqnarray}}
\def\ea{\end{eqnarray}}
\def\bea{\begin{eqnarray}}
\def\eea{\end{eqnarray}}
\def\eq{\begin{equation}}
\def\eqe{\end{equation}}
\def\eqa{\begin{eqnarray}}
\def\eqae{\end{eqnarray}}

\def\beqa{\begin{eqnarray}}
\def\eeqa{\end{eqnarray}}

\newcommand{\beqas}{\begin{eqnarray*}}

\newcommand{\eeqas}{\end{eqnarray*}}

\title{Exact coefficients for higher dimensional operators with sixteen supersymmetries}
\author[a]{Wei-Ming Chen}
\author[a,b]{ Yu-tin Huang}
\author[c]{Congkao Wen}
\affiliation[a]{Department of Physics and Astronomy, National Taiwan University, Taipei 10617, Taiwan, ROC} 
\affiliation[b]{School of Natural Sciences, Institute for Advanced
Study, Princeton, NJ 08540, USA}
\affiliation[c]{ I.N.F.N. Sezione di Roma ``Tor Vergata", Via della Ricerca Scientifica, 00133 Roma, Italy}
\emailAdd{tainist@gmail.com, yutinyt@gmail.com, Congkao.Wen@roma2.infn.it} 

\abstract{We consider constraints on higher-dimensional operators for supersymmetric effective field theories. In four dimensions with maximal supersymmetry and SU(4) R-symmetry, we demonstrate that the coefficients of abelian operators $F^n$ with MHV helicity configurations must satisfy a recursion relation, and are completely determined by that of $F^4$. As the $F^4$ coefficient is known to be one-loop exact, this allows us to derive exact coefficients for all such operators. We also argue that the results are consistent with the SL(2,Z) duality symmetry. Breaking SU(4) to Sp(4), in anticipation for the Coulomb branch effective action, we again find an infinite class of operators whose coefficients are determined exactly. We also consider three-dimensional $\mathcal{N}=8$ as well as six-dimensional $\mathcal{N}=(2,0),(1,0)$ and $(1,1)$ theories. In all cases, we demonstrate that the coefficient of dimension-six operator must be proportional to the square of that of dimension-four. }

\begin{document}
\begin{large}

\maketitle 
%%%%%%%%%%%%%%%%%%%%%%%%%%%%%%%%%%%%%%%%%%%%%%%%%%%%%%%%%%%%%
\section{Introduction and motivations}
%%%%%%%%%%%%%%%%%%%%%%%%%%%%%%%%%%%%%%%%%%%%%%%%%%%%%%%%%%%%%

The dynamics of the lower energy effective theory are encoded in the coefficients of higher-dimensional operators, which corresponds to polynomial expansions of constant field strengths and derivative expansions thereof. In principle these coefficients can be obtained by integrating away the massive degrees of freedom in the path integral. However this is difficult to preform exactly, and moreover the underlying Lagrangian may not even be known. On the other hand it is long known that supersymmetry imposes non-trivial constraints, and in some cases, the coefficients can be determined exactly. Early examples for four-dimensional supersymmetric gauge theories are the work of Dine and Seiberg~\cite{DineSeiberg}, which determines four derivative terms (include $F^4$) exactly by using half-maximal supersymmetry and conformal symmetry. Maximal supersymmetry without conformal symmetry can also determine four derivative terms in one dimension~\cite{PSSF41D}, three dimensions~\cite{PSSF43D} as well as theories with finite $N$~\cite{PSSN}. Furthermore, for $\mathcal{N}=4$ quantum mechanics, one can show that the coefficient of six-derivative terms are completely determined by that of four derivatives~\cite{PSSF61D}. Supersymmetry has also been extensively used to study higher derivative terms in the effective actions of maximal supersymmetric gravity theories with many interesting exact results have been obtained~\cite{Green:1997tv, Berkovits:1997pj, Green:1998by, Pioline:1998mn, Berkovits:1998ex,  Green:2005ba, Basu:2008cf}.

The difficulty in going beyond six-derivative terms stems from the complication of determining the necessary deformations to the SUSY transformations, as well as ambiguities associated with field redefinitions in the effective action. On the other hand, similar difficulty was encountered in  determining of local counter terms for supergravity theories. There an alternative approach was developed by considering on-shell matrix elements associated with the local operator~\cite{ElvangCT}. The advantage of this approach is that SUSY is linearly realized regardless of the multiplicity~\cite{ElvangSUSYW}, where the information of nonlinear transformation rules, as well as the non-linear gauge symmetries, are encoded in the locality constraints for the matrix elements, i.e. that they can only have physical poles and the residues must factorize into lower-point matrix elements. Such an approach was recently extended to effective gravitational theories with maximal supersymmetry in diverse dimensions~\cite{XYSG, XYSG1}, as well as gauge theories~\cite{XYSY, XY20}.

In four dimensions, a possible non-renormalization theorem for abelian $F^n$ was conjectured in~\cite{Tseytlin}. In particular, explicit perturbative computations at two loops (the one-loop contribution to $F^6$ vanishes) showed that the coefficient of $F^6$ coincides with the effective action of a single D3-brane in the $AdS_5 \times S^5$ background.\footnote{Recently it was conjectured~\cite{Schwarz:2013wra} that the effective action of a single D3-brane in the $AdS_5 \times S^5$ background with one flux gives the full effective action of $\mathcal{N}=4$ SYM with SU(2) gauge group in the Coulomb branch.} Motivated by the $AdS$/CFT duality of $\mathcal{N}=4$ SYM and IIB string theory in $AdS_5 \times S^5$~\cite{AdSCFT}, it was argued that this coefficient for $F^6$ is two-loop exact. This is in agreement with the analysis of~\cite{Kuzenko:2004} where it was shown that the coefficient of $F^6$ must be proportional to the square of that of $F^4$, which is one loop exact. It was further conjectured in~\cite{Tseytlin} that for more general operator $F^{2 l+2}$, there is one and only one particular Lorentz structure for each $F^{2 l+2}$ which should have a ``protected" coefficient and
receive contributions only from the $l$-th loop order\footnote{This particular part of $F^{2 l+2}$ was conjectured to match onto the corresponding structure in the expansion of Dirac-Born-Infeld (DBI) action~\cite{Tseytlin}. As we will see that this statement is not precisely correct since only the so-called ``MHV" parts of DBI action are protected, while all the non-MHV operators are expected to receive all order corrections.}. 

In this paper, in a remarkable simple way we will prove the above conjectured non-renormalization theorems, and make it precise which particular part of the operator $F^{2 l+2}$ is protected. In particular, the field strength $F_{\mu \nu}$ can be separated into $(1,0)$ and $(0,1)$ representation of SL(2,C), corresponding to self-dual and anti self-dual field strengths respectively, which we will denote as $F_+$ and $F_-$. Thus a general operator $(F)^{2k}$ may be separated into a sum of operators of the form $(F_-)^{2p}(F_+)^{2q}$ with $p+q=k$, and we denote its coefficient as $c_0^{p,q}$ where the subscript $0$ indicates that it has no derivative in contract to the operator $D^mF^n$. We will prove that the coefficients of so-called maximally helicity violating (MHV) operators $(F_-)^{2}(F_+)^{2q}$ (as well as their parity conjugate $(F_+)^{2}(F_-)^{2q}$) are given by:\footnote{In this paper, the coefficients are defined up to an over all $\frac{1}{g^2_{YM}}$ in front of the action. Thus the coefficient of $F^2$ is $-\frac{1}{4}$.}
\eq
c_0^{1,q}=4^{q-1}(c_0^{1,1})^{q}\,,
\eqe
i.e.  they are determined in terms of that of the four-point operator $(F_-)^{2}(F_+)^{2}$. As the latter is known to be one-loop exact~\cite{DineSeiberg}, this implies that the operator $(F_-)^{2}(F_+)^{2q}$ receives contribution at $q$ loops and is exact. Furthermore, as there are no instanton corrections to $c_0^{1,1}$, this result also predicts the absence of instanton corrections for these set of operators. This is consistent with explicit one-instanton computations, which shows that such operators are absent in the one-instanton effective action~\cite{Workinprogress}. This result is obtained by showing that no $\mathcal{N}=4$ supersymmetric local matrix elements with SU(4) R-symmetry exist for such operators. This fact implies that the contribution of the local operator to the on-shell matrix element must cancel against polynomial terms generated through factorization diagrams involving lower multiplicity operators. This sets up a recursive construction that iteratively relates the coefficients of higher multiplicity operators to that of the leading higher-dimensional operator.

On the Coulomb branch, the SU(4) R-symmetry is expected to be broken to Sp(4). We also proceed to analyze classes of Sp(4) invariant operators. We find that similar to MHV operators, the coefficient of $(F_-)^{2}(F_+)^{2q} \phi$, denoted as $c_0^{1,q}(\phi)$, is again completely determined by the coefficient of $(F_-)^{2}(F_+)^{2} \phi$ and $(F_-)^{2}(F_+)^{2}$:
\eq\label{Result1}
c_0^{1,q}(\phi)=q4^{q-1}(c_0^{1,1})^{q-1}c_0^{1,1}(\phi)\,.
\eqe
The operator $(F_-)^{2}(F_+)^{2} \phi$ is only generated at one loop, and since $c_0^{1,1}$ is also one-loop exact, eq.(\ref{Result1}) yields the exact coefficient of $(F_-)^{2}(F_+)^{2q} \phi$ on the Coulomb branch. A similar statement is found for the operator $(F_-)^{2}(F_+)^{2} \phi^m$. We also explicitly compute the coefficient of $(F_-)^{2}(F_+)^{2} \phi$ and show that it is simply twice of that of $(F_-)^{2}(F_+)^{2}$. This result combined with eq.(\ref{Result1}), allows us to conclude that the exact effective action must contain: 
\eq
\sum_{q=1}^{\infty}(4)^{q-1}\left(-\frac{\lambda}{2(4\pi)^2}\right)^q\frac{(F_-)^{2}(F_+)^{2q}}{|X^2|^{2q}}\,,
\eqe
where $\lambda=N*g^2_{YM}$, and $X^2$ is the SO(6) invariant inner product of the six scalars. Note that while this result coincides exactly with that of DBI action in $AdS_5 \times S^5$ background, and it is valid for all $N$.

For more general non-MHV operators as well as any $D^m F^n$ operators with $m \geq 4$, there exists local supersymmetric matrix elements and hence their coefficients are not tied to other lower-point operators. We believe that such operators generally receive all-loop as well as instanton corrections. Indeed from an one-loop general computation in~\cite{Tseytlin}, one can find that, unlike the MHV operators, all the non-MHV operators $(F_-)^{2p}(F_+)^{2q}$ start to appear already at one loop. Furthermore, perturbative loop and instanton computation shows that $D^4 F^4$ is not protected either~\cite{Workinprogress}.

We extend our analysis to theories in three dimensions and six dimensions. Unlike four dimensions, the R-symmetry of these theories generically contains a U(1) subgroup whose generator enforces uniform degree of Grassmann parameters for a given multiplicity, and thus there exists no similar helicity categorization as in four dimensions. Instead, we will only focus on dimension-six operators. We find that for the theory with maximal $\mathcal{N} =8$ supersymmetry in three dimensions, there is no SUSY completion, and we deduce that the coefficient of the dimension-six operator must be proportional to the square of dimension-four operator in the theory. The precise coefficient can be read off from the three-dimensional DBI action. This result applies to theories with SO(8) and SO(7) R-symmetry, where the latter corresponds to that of SYM. Note that it is known that the dimension-four operator receives perturbative and non-perturbative corrections~\cite{DineSeiberg, PSSF43D}. This result immediately yields the corresponding corrections for the dimension-six operator. We find the same conclusion for $\mathcal{N} =(2,0)$, $\mathcal{N} =(1,0)$ and $\mathcal{N}=(1,1)$ theories in six dimensions. \footnote{The absence of local dimension-six operators for $\mathcal{N} =(2,0)$ was already noticed in~\cite{Maxfield}.}

The paper is organized as follows: In section \ref{4D}, we first introduce the general idea of our approach, by studying the SUSY completion of the S-matrix elements associated with the local operators of our interest, for particular class of operators we can make precise statements which relate the coefficients of higher-point operators to that of lowest-point operator. We begin with four-dimensional theories with maximal supersymmetry. We find there is no SUSY completion for the S-matrix of the MHV operator $(F_{-})^2 (F_+)^{2q}$ as well as that of the SU(4) breaking operator $(F_{-})^2 (F_+)^{2q} \phi$, and thus lead to recursion relations for the coefficients of these two classes of operators, with recursions given in equations (\ref{Predict1}) and (\ref{Predict2}). We also comment that our findings are consistent with SL(2,Z) symmetry of $\mathcal{N}=4$ SYM. We then move on to theories in other dimensions, unlike in the case of four dimensions where one can classify the operators by their helicity configurations, in three and six dimensions we only consider dimension-six operators. In section \ref{3D}, we study the S-matrix of the dimension-six operator in a three-dimensional theory with maximal supersymmetry. We again find such S-matrix cannot exist to be consistent with $\mathcal{N}=8$ supersymmetry in three dimensions, and thus we conclude that the coefficient of the dimension-six operator must be proportional to the square of the coefficient of the dimension-four operator, and for the later the result is known perturbatively and non-perturbatively. In section \ref{6D}, we extend our analysis for theories in six dimensions with various choices of supersymmetry: $(2,0)$, $(1,0)$ and $(1,1)$. We find for all these cases, there is no consistent supersymmetric S-matrix with right properties, and thus the same as the three-dimensional case, the coefficients of dimension-six operators in these theories are all determined by those of dimension-four operators. We finish the paper with conclusions and remarks in section \ref{conclusion}.

%%%%%%%%%%%%%%%%%%%%%%%%%%%%%%%%%%%%%%%%%%%%%%%%%%%%%%%%%%%%%%%%%%%%%%%%%%
\section{Four dimensions\label{4D}}
%%%%%%%%%%%%%%%%%%%%%%%%%%%%%%%%%%%%%%%%%%%%%%%%%%%%%%%%%%%%%%%%%%%%%%%%%%

In four dimensions, with the aid of spinor helicity formalism and helicity decomposition, it is possible to determine the absence of local supersymmetric invariant matrix elements for a large class of operators. Here we will follow the approach developed for counter terms of $\mathcal{N}=8$ supergravity by Elvang, Freedman and Kiermaier~\cite{ElvangCT}. The absence of local SUSY matrix elements implies that factorization diagrams will produce local polynomials that exactly cancels the corresponding bosonic operator. As the factorization diagram involves lower multiplicity operators, SUSY fixes the coefficient of the operator in question in terms of lower multiplicity ones. Thus the coefficient of any local operator that does not have a corresponding local super-matrix element  is fixed in terms of lower multiplicity ones, perturbatively and non-perturbatively.

%%%%%%%%%%%%%%%%%%%%%%%%%%%%%%%%%%%%%%%%%%%%%%%%%%%%%%%%%%%%%%%%%%%%%%%%%%
\subsection{Local SUSY invariants with maximal supersymmetry}
%%%%%%%%%%%%%%%%%%%%%%%%%%%%%%%%%%%%%%%%%%%%%%%%%%%%%%%%%%%%%%%%%%%%%%%%%%
We begin by identifying which operators' matrix elements do not have local SUSY completion. The analysis is a two-step process: 
\begin{itemize}
  \item First construct the most general invariant respecting supersymmetry and R-symmetry. This determines the polynomial dependence on the grassmann variables, up to pure kinematic factors. The kinematic factors are fixed in terms of a few component matrix elements, by projecting the latter out from the super-function. The number of basis elements needed can be greatly reduced by employing permutation symmetry for abelian theories we are considering. This yields the supersymmetric completion of the basis component elements. 
  
  \item Apply multi-line shifts~\cite{Risager:2005vk} to probe the singularities of the super-function, which generically have manifest poles. If one can show that the poles do not cancel for any one of the component elements, then there exists no local SUSY completion.
  \end{itemize}

We will briefly review the process using $\mathcal{N}=4$ SU(4) R-symmetry as an example. Due to the SU(4) R-symmetry, matrix elements must be of degree $4(k{+}2)$ polynomials in $\eta^I$, which correspond to N$^{k}$MHV elements. 

\noindent \textbf{MHV matrix elements}:

We begin with the MHV case, for which the super-matrix element that reproduces the correct component amplitude $A_{n}(-,-,+,\cdots,+)$ is written as:
\eq
\mathcal{A}_{n}=\frac{\delta^8(Q)}{\langle12\rangle^4}A_{n}(-,-,+,\cdots,+)\,,
\eqe 
where the supercharge conservation is defined as 
\eq
\delta^8(Q)= \prod^2_{\alpha=1} \prod^4_{I=1} \left( \sum_{i}\lambda^{\alpha}_i\eta_i^I \right) \, .
\eqe
Here we use the standard spinor-helicity formalism,
\eq
p^{\alpha \dot{\alpha}}_i = \lambda^{\alpha}_i \tilde{\lambda}^{\dot{\alpha}}_i \,,
\eqe 
and scalar products are given by 
\eq
\lambda_i^\alpha\lambda_j^\beta\epsilon_{\alpha\beta}=\langle ij\rangle \,, \quad \tilde\lambda_{i\dot\alpha}\tilde\lambda_{j\dot\beta}\epsilon^{\dot\alpha\dot\beta}=[ij] \,, \quad 
s_{ij}=\langle ij\rangle[ji] \, .  
\eqe

We will now check if all component amplitudes coming from the above is local. To test locality, we will perform three-line shifts on legs $1,2$ and $n$,\footnote{A particular solution for $c_i$s that satisfy the last constraint, which is necessary for momentum conservation, is given by $c_1=[2n]$, $c_2=[n1]$ and $c_n=[12]$.}
\eq
\lambda_i \rightarrow \lambda_{\hat{i}} = \lambda_i + z c_i \xi \, ,  \quad {\rm with} \quad 
\sum_{i=1,2,n} c_i \tilde{\lambda}_i = 0 \, .
\eqe
This deforms the matrix element into a function of complex variable $z$, which allows one to more readily study the pole structure of the function in question. Note that we've chosen the shift such that the  $\langle12\rangle$ pole has been deformed, and since we expect no poles in $A_{n}$ as they are the matrix elements of some local bosonic operator, this ensures that all possible singularities of the super-function has been detected. 

Finally, let us project out a component amplitude that is not the original basis element, say $A_{n}(+,+,-,-,+,\cdots,+)$, which is given by 
\eq\label{NonLocal}
A_{n}(\hat{+},\hat{+},-,-,+,\cdots,\hat{+})=\frac{\langle 34\rangle^4}{\langle \hat{1}\hat{2}\rangle^4}A_{n}(\hat{-},\hat{-},+,\cdots,\hat{+})\,,
\eqe
and we have done a three-line shift on legs $n, 1$ and $2$. The question at hand is whether or not the fourth power pole may be cancelled by deformed spinor brackets in $A_{n}(\hat{-},\hat{-},+,\cdots,\hat{+})$. This can be determined by the mass dimension of the operator in question.

Let us first consider abelian $F^n$ operators. In four dimensions, one can write the field strength as:
\eq
F_{\mu\nu}\rightarrow F_{\alpha\dot{\alpha},\beta\dot{\beta}}=\epsilon_{\alpha\beta}(F_+)_{\dot\alpha\dot\beta}+\epsilon_{\dot\alpha\dot\beta}(F_-)_{\alpha\beta}
\eqe
where $F_-$ and $F_+$ are the self-dual and anti-self-dual field strengths respectively. In this language an $F^n$ operator can be decomposed into a sum of polynomials of the form $(F_-^{2})^p(F_+^{2})^q$ with $p+q=n/2$, and $F_-^{2}, F_+^2$ indicates their SL(2,C) indices are contracted. Note that here, the separation into $F_+$ and $F_-$ is simply due to irreducible representations of SL(2,C) and thus holds as an off-shell statement. Of course when on-shell, they naturally settle into positive and negative helicity states respectively.

For MHV due to its helicity configuration as well as the mass-dimension constraint, the basis element in eq.(\ref{NonLocal}) must have 2 $\lambda_1$'s, 2 $\lambda_2$'s, and 2 $\tilde\lambda_k$'s for $k \neq 1,2$. In other words, we can only have 2 spinor brackets in the numerator, and thus is not sufficient to cancel the fourth order pole. \textit{Thus MHV local matrix elements do not exist for $F^n$ operators for $n>4$}. For $n=4$, we have shifted numerator in $\langle 3\hat{4}\rangle$, and momentum conservation allows cancellation to yield a local polynomial.

Next let's consider $D^{2p}F^n$ operators, again for the case of MHV, it requires 2 $\lambda_1$'s, 2 $\lambda_2$'s, and 2 $\tilde\lambda_k$'s for $k \neq 1,2$, but now with $2p$ additional pairs of $|j\rangle[j|$ where $j$ can be arbitrary legs. This implies $2+p$ spinor brackets, and thus can only cancel the poles if $p>1$, i.e. $D^2F^n$ operators do not have a SUSY local completion for MHV matrix elements. Let us see if $D^4F^6$ exists. Now the requirement is that 
\eq
\frac{A_{n}(-,-,+,+,+,+)}{\langle12\rangle^4}
\eqe 
to be local. This implies that it must be (here perm indicates permutation in $\{3,4,5,6\}$ )
\eq
\left([12]^2[34]^2[56]^2+{\mathrm{perm} }\right)
\eqe
All other possibilities are equivalent via Schouten identities. Thus in conclusion, we have ruled out $F^n$ and $D^2F^n$ matrix elements as having MHV local matrix elements, but the S-matrix of $D^mF^n$ with $m \geq 4$ has a valid SUSY completion.

%%%%%%%%%%%%%%%%%%%%%%%
\noindent \textbf{NMHV matrix elements}:
%%%%%%%%%%%%%%%%%%%%%%%%%%
As discussed in~\cite{ElvangSUSYW}, for an N$^k$MHV matrix element, the amplitude is given by 
\eq
\mathcal{A}_n=\frac{\delta^8(Q)}{\langle nn{-}1\rangle^4}\sum_{I} c_{I}\prod_{a=1}^k(X_a)_{(i_a)_1,(i_a)_2,(i_a)_3,(i_a)_4}
\eqe
where $I$ labels the distinct Young tableaux where the indices  $\{(i_a)_1,(i_a)_2,(i_a)_3,(i_a)_4\}$ for a given $a$ populates a row and are hence symmetrize while each column is antisymmetrized. The function $X$ is defined as:  
\eq
(X_a)_{(i_a)_1,(i_a)_2,(i_a)_3,(i_a)_4}=\prod_{A=1}^4\frac{\eta^A_{i_A}[n{-}2,n{-}3]+\eta^A_{n{-}2}[n{-}3,i_A]+\eta^A_{n{-}3}[i_A,n{-}2]}{[n{-}2,n{-}3]^4}
\eqe
and $c_I$s are component amplitudes with legs $n{-}3$, $n{-}2$, $n{-}1$, $n$ taking helicity $(+,+,-,-)$ respectively, while the remaining legs are determined by the $4k$ set of indices $\{(i_a)_1$,$(i_a)_2$, $(i_a)_3$, $(i_a)_4\}$. 

The important point is the presence of the fourth order pole $\langle nn{-}1\rangle$. To test whether the singularity of this pole is a true singularity, we perform a three-line shift on legs  $n{-}3$, $n{-}2$, and $n{-}1$, such that $\langle nn{-}1\rangle\rightarrow \langle n \widehat{n{-}1}\rangle$. Next, we project out the component matrix element $A_n(-,-,\cdots,+,+,+,-)$. Note that this component amplitude was chosen such that we have plus helicity on legs $n{-}3$, $n{-}2$, and $n{-}1$. This is advantageous as this isolates the shifted angle brackets to be completely contained in $c_{I}$, which are the component amplitudes. 

For NMHV $D^{2p}F^n$ operators, we must have 6 $\lambda_i$s and $(n{-}3)$ $\tilde{\lambda}_i$s, with additional $p$ angle and square brackets respectively. Thus besides $p=0$, we will in principle have enough angle brackets to cancel the fourth order pole. However, for $p=0$, the matrix element is proportional to $3$ angle brackets constructed from the three negative helicity legs, and vanishes under permutation symmetry. One might wonder if other component operators can have non-zero matrix elements, and thus we would have a non-local SUSY element. In appendix~\ref{SUSYWard}, we will show that using permutation symmetry, we can express all possible components in terms of that of $F^n$ for $n=6$, and thus if the S-matrix of $F^6$ vanishes, so must each the whole super-matrix element. Thus there exists no local super-matrix elements for NMHV with mass-dimensions $n$, at least for $n=6$. For any non-zero $p$'s, there can be NMHV local amplitudes. For example, an explicit form of local SUSY completion of $D^2F^6$ was given in ~\cite{ElvangCT}. Beyond NMHV, in general there will be sufficient number of angle brackets to cancel the higher order poles.

It is intriguing to see what conclusions one can draw with reduced supersymmetry. Such scenario may arise if one considers corrections due to BPS objects. For $\mathcal{N}=2$ and $\mathcal{N}=1$, the on-shell degrees of freedom are encapsulated in two distinct superfields,  $\Phi,\,\Psi$, which contains the positive and negative helicity vector respectively~\cite{ElvangH}. Thus there are distinct $\frac{n!}{(k{+}2)!(n{-}k{-}2)!}$ super-matrix elements for N$^2$MHV configuration. For MHV, the super-matrix element that contains $m(-,-,+,\cdots,+)$ for $F^n$ can be written as:
\eq
\mathcal{A}_n=\frac{\delta^{2\mathcal{N}}(Q)}{\langle12\rangle^{\mathcal{N}}}A_n(-,-,+,\cdots,+)\,,
\eqe
where $\mathcal{N}=1,2$. Following the previous analysis we can see that local super-matrix elements \textit{can} exists for $\mathcal{N}=1,2$. For example, one has:
\eqa
\nonumber\mathcal{N}=2&:& \; \delta^{4}(Q)([34]^2[56]^2\cdots[nn{-}1]^2{+}perm)\,\\
\quad \mathcal{N}=1&:& \; \delta^{2}(Q)\langle12\rangle([34]^2[56]^2\cdots[nn{-}1]^2{+}perm)\,.
\eqae
Note that for $\mathcal{N}=1$ the super-matrix element is anti-symmetric in $1,2$. This is valid since $\Psi$ is fermionic for $\mathcal{N}=1$.

%%%%%%%%%%%%%%%%%%%%%%%%%%%%%%%%%%%%%%%%%%%%%%%
\subsection{Implications for non-renormalization theorems}
%%%%%%%%%%%%%%%%%%%%%%%%%%%%%%%%%%%%%%%%%%%%%%%
We now consider the implications of the above results for the higher dimensional operators in an effective theory with $\mathcal{N}=4$ supersymmetry and SU(4) R-symmetry. In the constant back ground approximation, we can write an effective Lagrangian as:
\eqa
\label{efL}\mathcal{L}_{\rm eff}
&=&\sum_{p,q=1} c_0^{p,q}\frac{(F^2_+)^p(F^2_{-})^q}{(M^2)^{2(p+q-1)}}+\sum_{m=1}\sum_{p,q=1} c_m^{p,q}\frac{D^{2m}(F^2_+)^p(F^2_{-})^q}{(M^2)^{2(p+q-1)+m}}+ \cdots
\eqae
where $\cdots$ indicate its possible SUSY completions and $M$ is some UV cut-off. In general there are no constraints on $c_m^{p,q}$, and in the process of integrating away the massive degrees of freedom, they can receive all-loop perturbative as well as non-perturbative contributions. As we have seen, $\mathcal{N}=4$ supersymmetry with SU(4) R-symmetry tells us that local matrix elements for MHV. helicity configurations do not exists for $F^n$ operators. Note that the lack of MHV local supersymmetric matrix element implies that the MHV amplitude is simply zero. This is because for an abelian U(1) theory, the MHV amplitude cannot have any factorization poles. Thus the only possibility \textit{is} a local polynomial which we've just shown to not exist.

Furthermore, this also relates the coefficients of higher-point operators $c_0^{1,p}$ to lower-point ones. For example at six points two $F^4$ operators must generate exactly the same local polynomial as $(F_-)^2(F_+)^4$, and thus the coefficient of the latter must be the opposite of the coefficient of this polynomial. This is indeed exemplified by the DBI action as discussed in~\cite{Rosly:2002jt, Rutger}. We will show that this will allow us to determine the exact coefficient for MHV operators with arbitrary multiplicity.

In the Feynman-'t Hooft gauge, the distinct field strengths contract as
\eq\label{Props}
(F_{-})_{\alpha\beta}(F_{-})_{\gamma\delta}\rightarrow i(\epsilon_{\alpha \gamma}\epsilon_{\beta\delta}+\epsilon_{\alpha \delta}\epsilon_{\beta\gamma})\,,\quad (F_{+})_{\dot{\alpha}\dot{\beta}}(F_{-})_{\gamma\delta}\rightarrow -\frac{i(p_{\alpha\dot{\gamma}}p_{\beta\dot{\delta}}+p_{\beta\dot{\gamma}}p_{\alpha\dot{\delta}})}{p^2}\,,
\eqe 
and the Feynman rules for $(2n)$-point MHV operator $c_0^{1,n{-}1}(F^-)^2[(F^+)^2]^{n{-}1}$ in the effective Lagrangian eq.\eqref{efL} is 
\eq\label{FV}\hspace{-0.95cm}
\begin{array}{c}
\includegraphics[scale=1]{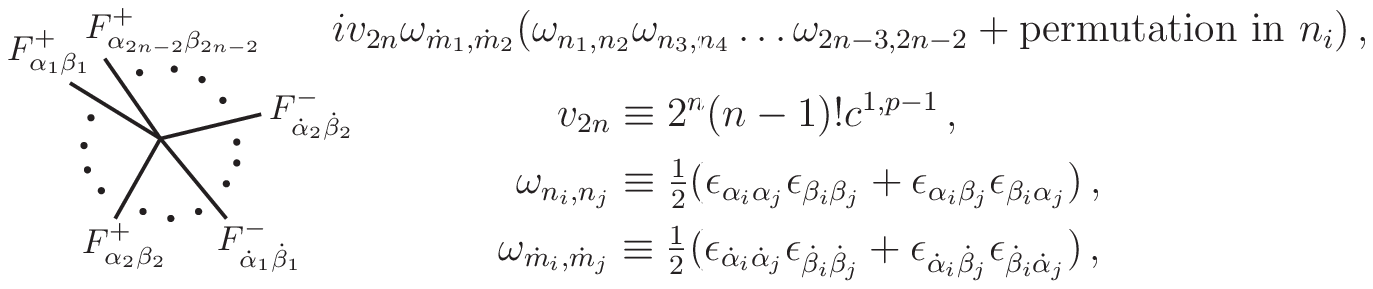}
\end{array}\,.
\eqe
where it is understood that the free indices are to be contracted with external line factors of the field strength, i.e. $\lambda^{\alpha}_i \lambda^{\beta}_i$ and $\tilde{\lambda}^{\dot{\alpha}}_i\tilde{\lambda}^{\dot{\beta}}_i$.

Let us consider the six-point matrix element with $(F^-_{i}\cdot F^-_{j})(F^+_{k}\cdot F^+_{l})(F^+_{m}\cdot F^+_{n})$. From the Feynman rules one can deduce that only diagrams where legs ($k$, $l$) and ($m$, $n$) sit at the same vertex respectively will contribute. Thus the counting amounts to how many ways one can distribute the pairs of plus helicity field strength across a graph. At six-point we have:
\eq
\label{blackL}\begin{array}{c}
\includegraphics[scale=1]{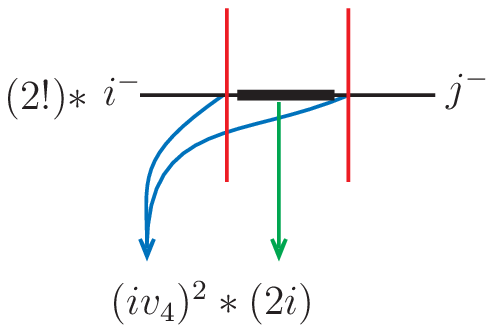}
\end{array}
\eqe
where in the above diagram the red line are with the external field strength $F^+$s, and the thick black line represents the contraction of two vertices. In this example, one learns that: (1) each quartic diagram with negative helicities contracted with each other yields a polynomial identical to a term in $(F_-)^2(F_+)^4$. (2) Each contraction yields a factor of $2i$ from propagator. (3) the $2!$ factor in eq.\eqref{blackL} is the number of ways the plus helicity pairs can be distributed across the diagram. Thus the coefficient of $v_6$ must be the opposite of this counting.

In fact, it will be convenient to blow up the six-point vertex as a factorization diagram with a ``wrong sign" propagator:  
\eq
\begin{array}{c}
\label{greyL}\includegraphics[scale=1]{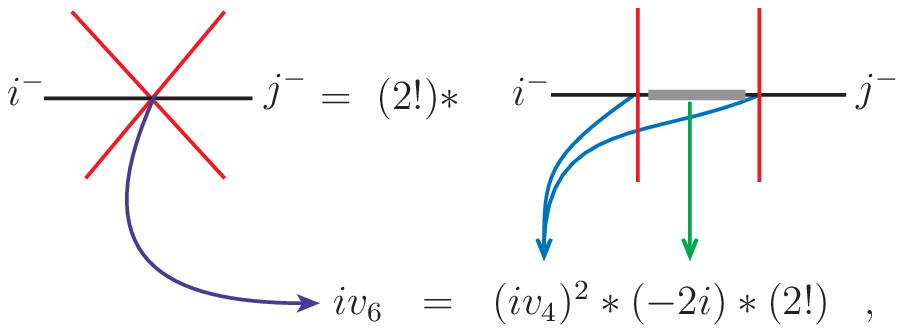}
\end{array}
\eqe
where the grey line represents the contraction between two vertices which gives a factor $-2i$ in matrix elements (in contrast to the factor $2i$ represented by the black line in eq.\eqref{blackL}). We will see that the notation of black and grey lines for contractions will be useful in proving general case. We can study 8-point matrix element to understand the general pattern between the vertex coefficients. In the 8-point case, diagrammatically, we have
\eq
\begin{array}{c}
\hspace{-0.6cm}\includegraphics[scale=1]{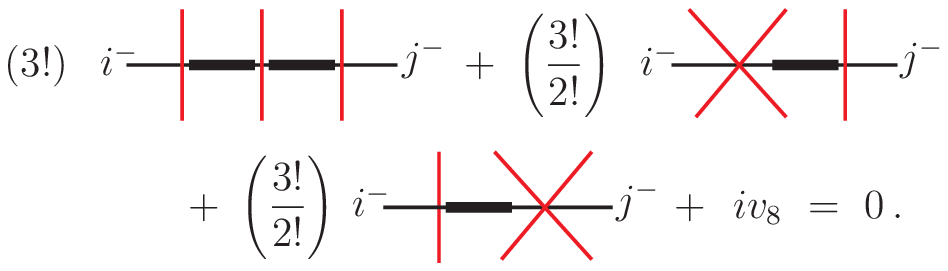}
\end{array}
\eqe
We can use eq.\eqref{greyL} to express six-point vertex as two four-point vertices joined by a grey line 
\eq
\begin{array}{c}
\vspace{-3mm}\hspace{-0.4cm}\includegraphics[scale=1]{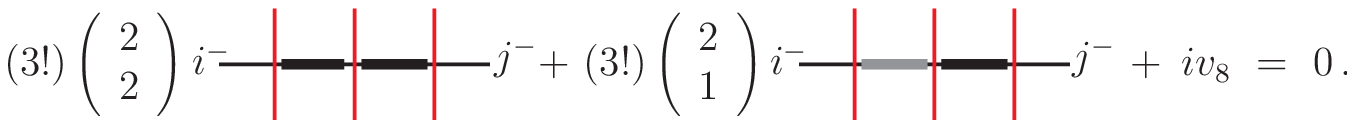}
\end{array}
\eqe
One can see that the coefficient of 8-point vertex is related to join three four point vertices with at least one black line and at most two black lines. The combinatorial factor in each diagram is exactly the number of ways one can have to join the four-point vertices with a given number of black and grey lines. Moreover, we know the black line gives a factor $2i$ and the grey line gives $-2i$, so we can replace $n$ black lines by $n$ grey lines with an additional factors $(-1)^n$, this yields
\eq
\begin{array}{c}
\vspace{-3mm}\label{bi8}\includegraphics[scale=1]{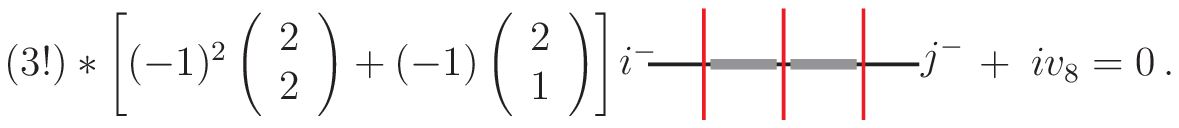}
\end{array}
\eqe
From binomial expansion, we know
\eq
\left(\begin{array}{c}
n\\
n
\end{array}\right)(-1)^n+\left(\begin{array}{c}
n\\
n-1
\end{array}\right)(-1)^{n-1}\dots+\left(\begin{array}{c}
n\\
1
\end{array}\right)(-1)=-1\,,
\eqe
thus we have
\eq
\begin{array}{c}
\vspace{-3mm}\includegraphics[scale=1]{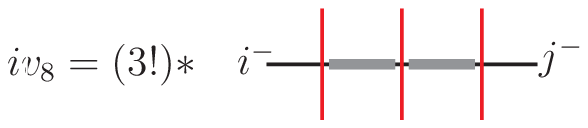}
\end{array}\,,
\eqe
which is similar to the result of six-point vertex coefficient and can be considered as three four-point vertices joined by two grey lines. In general case, for a $2n$-point vertex coefficient, we can divide all the diagrams into $(n-2)$ categories of diagrams with $1,\dots,j,\dots,(n{-}2)$ contractions with black lines. Moreover, because we can recast all vertices into four-point vertices with contractions by grey lines, a category with $j$ contractions by black lines contributes
\eq 
(n-1)!\left(\begin{array}{c}
n-2\\
j
\end{array}\right)(iv_4)^{n-1}
\underbrace{(2i)^j}_{\mathrm{contractions\atop{~by~black~lines}}}
 \underbrace{(-2i)^{n-2-2j}}_{\mathrm{contractions\atop{~by~grey~lines}}}\,.
\eqe
The sum of all categories again forms a binomial expansion like the one in eq.\eqref{bi8}, and we have
\eq 
iv_{2n}=(n-1)!(iv_4)^{n-1}(-2i)^{n-2}\,,
\eqe
or 
\eq\label{Predict1}
c_0^{1,q}=4^{q-1}(c_0^{1,1})^{q}\,.
\eqe
So we completely determine the coefficient $c_0^{1,q}$ in terms of $c_0^{1,1}$. If explicit form for $c_0^{1,1}$ is known, we then have precise formula for any $c_0^{1,q}$.

This is precisely the case for $\mathcal{N}=4$ SYM in four dimensions. It was discussed in~\cite{Tseytlin}, where the $F^4$ operator is generated at one loop with $c^{1,1}_0=-\frac{\lambda}{2(4\pi)^2}$, and is non-renormalized through all loops as well as non-perturbatively. Thus combined with the above analysis, it gives the exact coefficient for $c_0^{1, q}$ for the effective action of $\mathcal{N}=4$ SYM on the Coulomb branch:
\eq\label{EffectiveResult}
\sum_{q=1}^{\infty}(4)^{q-1}\left(-\frac{\lambda}{2(4\pi)^2}\right)^q\frac{(F_-)^{2}(F_+)^{2q}}{|X^2|^{2q}}\,,
\eqe  
where $\lambda$ is the t'Hooft coupling $Ng^2_{YM}$ and $N$ correspond to SU(N). Note that this result is valid for any $N$, and is perturbatively and non-perturbatively exact. This is consistent with explicit computations with one-instanton corrections~\cite{Workinprogress}, the results show that indeed all the MHV operators $(F_-)^{2}(F_+)^{2q}$ are absent in the one-instanton effective action. The above result also tells us that $(F_-)^{2}(F_+)^{2q}$ is only generated at $q$-loop, with only planar sectors contributing, and do not receive any higher-loop corrections. In the following we will argue that the result is consistent with the known SL(2,Z) duality symmetry of $\mathcal{N}=4$ SYM.

%%%%%%%%%%%%%%%%%%%%%%%%%%%%%%%%%%%%%%%%%%%%%%%
\subsection{Modular invariance}
%%%%%%%%%%%%%%%%%%%%%%%%%%%%%%%%%%%%%%%%%%%%%%%
We have shown that maximal supersymmetry completely fixes the coefficient of $F^n$ MHV operators. An interesting question is whether or not this is consistent with the SL(2,Z) duality symmetry. This question can be made more precise by considering that these coefficients exactly coincide with DBI, which is known to be duality invariant~\cite{SDuality}. On the other hand, there is strong evidence to believe that the complete effective action on the Coulomb branch will contain additional terms beyond that of DBI. For example, an explicit computation has shown evidence for all-loop renormalization for N$^2$MHV $(F_-)^4(F_+)^4$ operators~\cite{Fradkin:1982kf, Chepelev:1997av}. The question then becomes whether or not such deformations are allowed under the constraint of exact coefficients for MHV operators and duality symmetry. Here we will demonstrate the answer is positive.

The SL(2,Z) duality transformations acts on the complexified coupling $\tau = {\theta \over 2 \pi } + {4 \pi i \over g^2_{\rm YM}} :=  \tau_1 + i \tau_2 \, ,$ as, 
\eqa
\tau \rightarrow { a \tau +b \over c \tau + d } \, ,
\eqae
where $a, b, c,d \in Z$ and $ad-bc=1$. The field strength $F_{\mu \nu}$ together with its dual $G^{\mu \nu} = i \delta \mathcal{L}/ \delta F_{\mu \nu}$ form an $SL(2,Z)$ doublet, namely
\eqa \label{Rules}
 \begin{pmatrix} 
*G  \\
F
\end{pmatrix}   
\rightarrow 
\begin{pmatrix} 
a & b \\
c & d 
\end{pmatrix}
 \begin{pmatrix} 
*G  \\
F
\end{pmatrix}\, ,
\eqae
where $*$ denotes the Hodge dual. 
To study duality transformation of actions, it is useful to identify covariant objects under SL(2,Z). For example, the gauge coupling $\tau_2$ transforms nicely as:
\eqa
\tau_2 \rightarrow { \tau_2 \over (c \tau + d) (c \bar{\tau} + d)  } \, .
\eqae
For the field strengths, one can define the following linear combinations that are duality covariant: 
\eqa \label{PlusFDef}
\mathcal{F}^+ = {1 \over i \tau_2} (\tau F- *G)\, , \quad 
\mathcal{F}^- = {1 \over i \tau_2} (\bar{\tau} F- *G) \, ,
\eqae
such that under  eq.(\ref{Rules}), it transforms as
\eqa
\mathcal{F}^+ \rightarrow (c \bar{\tau} +d) \mathcal{F}^+ \, , \quad
\mathcal{F}^- \rightarrow (c {\tau} +d) \mathcal{F}^- \, .
\eqae
If we take the limit where the theory becomes non-interacting, i.e. all higher dimension operators are set to zero, then $\mathcal{F}^{\pm}$ defined above are simply self-dual and anti-self-dual field strengths, ${F}^{\pm}$. Thus under SL(2,Z) transformations, ${F}^{\pm}$ transforms covariantly at this order. 

For the interacting case, the covariant field strengths $(\mathcal{F}^+,\mathcal{F}^-)$ can contain both self-dual and anti-self dual field strengths. They can be separated into SL(2,C) irreducible representations:
\eqa
\nonumber (\mathcal{F}^-)_{(\dot\alpha\dot\beta)}&=&(F^-)_{\dot\alpha\dot\beta}+(F^-)_{\dot\alpha\dot\beta}[b_1(F^+)^2+ b_2(F^-)^2]+\cdots\\
(\mathcal{F}^+)_{(\dot\alpha\dot\beta)}&=&0+\quad (F^-)_{\dot\alpha\dot\beta}[b_1(F^+)^2+ b_2(F^-)^2]+\cdots+\cdots\
\eqae
where $b_i$s are coefficients determined by $\delta S/\delta F_-$. Now while duality symmetry rotates between $(\mathcal{F}^-)_{(\dot\alpha\dot\beta)}$ and $(\mathcal{F}^+)_{(\dot\alpha\dot\beta)}$, one can re-express the transformation as a non-linear redefinition of the field strength. In particular, the negative field strength would transform as:
\eq
F^-\rightarrow \bar{\tau} F^-+\alpha F^-(F^+)^2+\cdots
\eqe
The important point is that under the duality transformation, the number of negative field strengths cannot decrease. Thus MHV operators only receives contribution from itself under duality transformation and is completely isolated from all the N$^k$MHV operators. Thus from the MHV operators point of view, any deformation of the N$^k$MHV operators do not participate in the duality transformation of eq.(\ref{EffectiveResult}).

%%%%%%%%%%%%%%%%%%%%%%%%%%%%%%%%%%%%%%%%%%%%%%%
\subsection{Sp(4) invariants}
%%%%%%%%%%%%%%%%%%%%%%%%%%%%%%%%%%%%%%%%%%%%%%%
On the Coulomb branch, prior to integrating away the massive modes, the R-symmetry of the theory is broken down from SU(4) to Sp(4). After integrating away the massive modes, we expect terms in the effective action that carries this SU(4) breaking fingerprint. Thus in this section we proceed and analyze general Sp(4) invariants.

Sp(4) generators $G_{IJ}$ can be obtained from linear combination of SU(4) ones $G_{I}\,^J$ as 
\eq
G_{IJ}=G_{(I}\,^K\Omega_{KJ)}
\eqe
where $\Omega_{IJ}$ is the Sp(4) metric. For us we will choose 
\eq
\Omega_{13}=-\Omega_{31}=-1,\;\;\Omega_{24}=-\Omega_{42}=-1
\eqe
all other entries are zero. As a result we have the following 10 generators,
\eqa\label{Gen}
G_{12}=G_{1}\,^4+G_{2}\,^3,\;\;G_{23}=-G_{2}\,^1+G_{3}\,^4,\;\;G_{34}=-G_{3}\,^2-G_{4}\,^1\nonumber\\
G_{41}=G_{4}\,^3-G_{1}\,^2,\;\;G_{13}=-G_{1}\,^1+G_{3}\,^3,\;\;G_{24}=-G_{2}\,^2+G_{4}\,^4\nonumber\\
G_{11}=G_{1}\,^3,\;\;G_{22}=G_{2}\,^4,\;\;G_{33}=-G_{1}\,^3,\;\;G_{44}=-G_{4}\,^2\,.
\eqae
Let's consider the solution to SUSY Ward identity with Sp(4) symmetry. First of all from eq.(\ref{Gen}) we see that the set of generators 
$(G_{11},G_{33},G_{13})$, and $(G_{22}, G_{44}, G_{24})$ forms an SU(2)$\times$SU(2) subgroup, while the remaining generators mixes the two. Thus we expect to write an Sp(4) invariant as a linear combination of the two SU(2) invariants. 

The first possible SU(4) breaking term is at five points.  Let's consider the minimum solution to these generators. Generally the superamplitude can be written as,
\eq
\mathcal{A}_{n}=\delta^8(Q)P_2, \quad P_2=\sum_{i,j,k,l=1}^n q_{ij}\eta_{i}^1\eta_j^3+p_{kl}\eta^2_k\eta_l^4\,,
\eqe
the subscript in $P_2$ indicates it's degree 2 polynomial in $\eta$s. The vanishing under $(G_{11},G_{33})$ and $(G_{22},G_{44})$ requires us to set $q_{ij}$ to be symmetric in $i,j$ and $p_{kl}$ to be symmetric in $k,l$. Invariance under $G_{12}$ requires
\eq
G_{12}\mathcal{A}_{n}=\delta^8(Q)\sum_{i,j,k,l=1}^n( q_{ij}\eta^1_i\eta_j^2+p_{kl}\eta^2_k\eta_l^1)=0
\eqe
This tells us that $q_{ij}=p_{kl}$. The same solution satisfies the remaining three generators, hence we arrive at: 
\eq
P_2=\sum_{i,j=1}^n q_{ij}(\eta^1_i\eta_j^3+\eta^2_i\eta_j^4)
\eqe
Following \cite{ElvangSUSYW}, the SUSY constraint imposed by $\bar{Q}^{\dot{\alpha}}_A= \sum_{i} \tilde{\lambda}^{\dot{\alpha}}_i \partial_{\eta^A_i}$ tells us that 
\eq
\mathcal{A}_{n}=\sum_{1\leq i\leq j\leq n-4}q_{ij}\delta^8(Q) \frac{(m_{i,n{-}3,n{-}2})_1(m_{j,n{-}3,n{-}2})_3+(m_{i,n{-}3,n{-}2})_2(m_{j,n{-}3,n{-}2})_4}{\langle n{-}1n\rangle^4[n{-}3,n{-}2]^2}\,,
\eqe
where 
\eq
(m_{i,n{-}3,n{-}2})_I=[n{-}3,n{-}2]\eta_{iI}+[n{-}2,i]\eta_{n{-}3I}+[i,n{-}3]\eta_{n{-}2I}
\eqe
From the helicity weight, we can see that 
\eq
q_{ij}=A_n(\{i,j\},+,+,-,-)
\eqe
where all unmarked legs are all of positive helicity gluons. Thus the total basis amplitudes contains a total of $(n{-}4)(n{-}3)/2$ of them. These are the minimal SU(4) breaking elements, whose bosonic component contain two minus helicity photons one scalar and arbitrary number of positive helicity photons. 

For simplicity let's define 
\eqa
Y_{i,j}=\frac{(m_{i,n{-}3,n{-}2})_1(m_{j,n{-}3,n{-}2})_3+(m_{i,n{-}3,n{-}2})_2(m_{j,n{-}3,n{-}2})_4}{[n{-}3,n{-}2]^2}\,.
\eqae
These will be the building blocks to generate general SU(4) breaking Sp(4) invariants. These will in general have degree $8+2k$ in $\eta$s with $k\in Z^+$, and we will refer them as N$^{\frac{k}{2}}$MHV matrix-elements. Let's consider some examples.

At five-point, we only need 1 and that is $A_n(\phi,+,+,-,-)$ that originates from the operator $\phi F^4$, the amplitude would be $[23]^2\langle 45\rangle^2$. Thus the superamplitude is just 
\eq\label{Sp45pt}
\mathcal{A}_5=\delta^8(Q) \frac{(m_{1,2,3})_1(m_{1,2,3})_3+(m_{1,2,3})_2(m_{1,2,3})_4}{\langle 45\rangle^2}
\eqe
Beyond five-point, one can have multiple factor of $Y$s. For example, at six-point a new solution to the Sp(4) SUSY Ward identity would be as follows:
 \eqa
\nonumber\mathcal{A}_{6}=\sum_{\footnotesize{\begin{array}{c} 1\leq i\leq j\leq 2 \\ 1\leq l\leq k\leq 2\end{array}}}\frac{\delta^8(Q)q_{ij,lk}}{\langle n{-}1n\rangle^4}Y_{i,j}Y_{l,k}
\eqae
The coefficient $q_{ij,lk}$ is symmetric in $(i,j)$ and $(l,k)$ while antisymmetric in $(i,l)$ and $(j,k)$. We have 
\eq
q_{ij,lk}=A_6\left(\left\{\begin{array}{c}i,j \\ l,k\end{array}\right\},+,+,-,-\right)
\eqe
The total basis amplitude is now 3 for six-point.

We now look at whether or not there are local supersymmetric elements for $(F_-)^2(F^+)^{2q}\phi^m$. We fist consider N$^{\frac{1}{2}}$MHV amplitudes. Note that since $(F_-)^2(F^+)^{2q}\phi^m$ have distinct mass-dimensions from $(F_-)^2(F^+)^{2q}(\bar{\psi}\psi)^\frac{m}{2}$, they have distinct SUSY completions. Thus for $(F_-)^2(F^+)^{2q}\phi$ we have
\eq
\mathcal{A}_{n}=\sum_{1\leq i\leq n-4}q_{i}\frac{\delta^8(Q)}{\langle n \, n{-}1\rangle^4}Y_{i}\,,
\eqe
with $n=2q+3$. Let us again perform a three-line shift on $n{-}2$, $n{-}1$, and $n{-}3$. We see that there is a fourth order pole in the denominator. Now let's consider projecting out A$_n(-,\cdots, \phi(i),+,+,+,-)$, this choice ensures that all angle brackets coming out of $\delta^8(Q)$ do not contain the shifted ones. Note that this choice is only possible for $n>5$ with N$^{\frac{1}{2}}$MHV amplitudes. In such case, the only possible angle brackets stem from $q_{i}$, which is simply the component matrix element $A_n(\cdots, \phi(i),+,+,-,-)$ for $(F_-)^2(F^+)^{2q}\phi$. This will have exactly 2 angle brackets, and thus will not be sufficient to cancel all the poles. For $n=5$ it is straight forward to see from eq.(\ref{Sp45pt}) that the poles cannot be canceled. Thus we conclude that there are no local $\mathcal{N}=4$ SUSY completion for $(F_-)^2(F^+)^{2q}\phi$. 

Note that the above result can be straightforwardly generalized to operators of $(F_-)^2(F^+)^{2q}\phi^m$. These simply correspond to matrix elements of the form 
\eq
\frac{\delta^8(Q)}{\langle n{-}1n\rangle^4}q_{i_{1},,\cdots,i_{m}}\prod_{a=1}^mY_{i_{1}}\cdots Y_{i_{a}}\,,
\eqe
where again the coefficients $q_{i_{1},\cdots,i_{m}}$ are given by linear combinations of component amplitudes $A_n(\cdots,\phi(i_1) ,\cdots,\phi(i_m),+,+,-,-)$ of $(F_-)^2(F^+)^{2q}\phi^m$. Since the latter still has two angle brackets and cannot cancel against the fourth order pole, this implies that its coefficient will again be iteratively determined by $(F_-)^2(F^+)^{2}\phi^m$, which in fact again is one-loop exact~\cite{DineSeiberg} and can be obtained from eq.(\ref{eq:DineSeiberg}). 

This proportionality can be easily fixed simply by considering the analogous diagrams as in SU(4) case, with summing over all possible insertions of extra scalar lines at each vertex. We will use $c_0^{1,q}(\phi)$ as an example. Beginning at seven points, we have two ways to assign the scalar line to one of the two vertices. As a result, relative to the six-point vertex in eq.(\ref{greyL}), there is an additional factor 2 in the seven-point vertex. Furthermore seven-point vertex can be viewed as a contraction of two four-point vertices with a overall factor 2 which is the number of possible ways for the scalar insertion. The procedure iteratively continues to higher-point vertices as what has been done before. In general, ($2n_v+3$)-point diagrams (except for the diagram from contact term) can be categorized into $n_v$ 4-point vertices with different numbers of contractions by black and grey lines and one scalar line assigned to the one of the $n_v$ vertices. In other words, for each category of ($2n_v+3$)-point diagrams, it produces $n_v$ number of scalar insertion diagrams. It turns out there is an overall $n_v$ factor for every category from the scalar insertion and the remaining contractions between 4-point vertices give the same result as in analysis of MHV matrix element. With $n_v=(n-3)/2$, this implies that 
\eq\label{Predict2}
c_0^{1,q}(\phi)=q \, 4^{q-1}(c_0^{1,1})^{q-1}c_0^{1,1}(\phi)\,.
\eqe
Since $c_0^{1,1}$ is protected and unrenormalized, this implies that $c_0^{1,q}(\phi)$ is completely determined by $c_0^{1,1}(\phi)$, which turns out is also one-loop exact as we will see shortly.

It is known that four derivative terms in $\mathcal{N}=4$ SYM given in the following expression is all-loop exact and does not receive any non-perturbative corrections~\cite{DineSeiberg}, 
\eq \label{eq:DineSeiberg}
S_4 \sim \int d^4 \theta d^4 \bar{\theta} \ln \left( { \Psi \over \Lambda }\right) \ln \left( { \bar{\Psi} \over \Lambda }\right) \, ,
\eqe
with $\Psi$ and $\bar{\Psi}$ are written in a $\mathcal{N}=2$ superspace, schematically,
\eq
\Psi = \phi + \sqrt{2} \lambda \theta + F_+ \theta^2 + \ldots \, , \quad 
\bar{\Psi} = \bar{\phi} + \sqrt{2} \bar{\lambda} \bar{\theta} + F_- \bar{\theta}^2 + \ldots \, .
\eqe
Now if we give a vev to the scalar field $\phi \rightarrow v + \phi$, and expand the action $S_4$ to fifth order we find the action contains, 
\eq
S_4 \ni {(F_+)^2 (F_-)^2 \over v^4} -2 \left[  {(F_+)^2 (F_-)^2 \phi \over v^5}+  {(F_+)^2 (F_-)^2 \bar{\phi} \over v^5} \right] \, .
\eqe
From this simple expansion we find the relative coefficient between operator $F^4$ and $F^4\phi$ is simply $-2$, that is $c_0^{1,1}(\phi)=-2c_0^{1,1}$, thus is also one-loop exact. This is consistent with expanding SO(6) invariant $F^4/|X \bar{X}|^2$ around a vev $v$. In the next section, we will explicitly compute the one-loop contribution to $c_0^{1,1}(\phi)$ and $c_0^{1,1}$ from the amplitude point of view.  

%%%%%%%%%%%%%%%%%%%%%%%%%%%%%%%%%%%%%%%%%%%%%%%%%%%%%%%%%%%%%%%%
\subsection{Explicit computation of $F^4\phi$ and $D^4F^4 \phi$}
%%%%%%%%%%%%%%%%%%%%%%%%%%%%%%%%%%%%%%%%%%%%%%%%%%%%%%%%%%%%%%%%
In this subsection, we will explicitly compute the one-loop coefficients for higher dimensional operators for $D^m F^4$ and $D^m F^4 \phi$. This is done by considering the one-loop amplitude with massive internal states, which can be obtain by dimensionally reducing the higher-dimensional integrand and identifying the extra dimension momenta to be the mass. Expanding around the large mass limit then correspond to integrating away the massive modes. In particular we will find that the ratio between $F^4$ and $F^4\phi$, as well as $D^4 F^4$ and $D^4 F^4 \phi$, is $-2$. 

The one-loop contribution to the operator $F^4\phi$ arises from the massive W boson multiplet circulating the loop, with one edge involving the massive multiplet coupling to the scalar that gets a vev~\cite{5thDim, ElvangMass}. The integrand of this amplitude can be obtained by considering the six-dimensional $\mathcal{N}=(1,1)$ integrand and the extra dimensional momenta are interpreted as the mass. More precisely,
\eq
\ell^2=(\ell^{(4D)})^2+\ell_4^2+\ell_5^2=(p^{(4D)})^2-m\tilde m\,,~~~~~~~m=\ell_4+i \ell_5\,,~\tilde m=\ell_4-i\ell_5\,.
\eqe
The five-point color-ordered one-loop amplitude was computed in~\cite{Brandhuber:2010mm} and the abelian U(1) amplitude is obtained by considering all possible distinguished permutations of external legs (and choosing the external fields as six-dimensional gluons):
\eq
\mathcal{A}_5=C^{\mu}\int d^6\ell \frac{\ell_\mu}{\ell^2(\ell+p_1)^2(\ell+p_1+p_2)^2(\ell+p_1+p_2+p_3)^2(\ell-p_5)^2}+(\mathrm{perm})
\eqe
with 
\eqa
C^\mu&=&\left[\frac{\langle 1_a2_b3_c4_d\rangle[1_{\dot{a}}2_{\dot{b}}3_{\dot{c}}4_{\dot{d}}][5_{\dot{e}}|p_1\sigma^\mu p_2p_1|5_e\rangle}{s_{34}s_{15}}+(1\leftrightarrow 2)+(5\leftrightarrow 1,\,1\leftrightarrow2)\right]\\
\notag&&
\hspace{-0.4cm}+
\left[\frac{\langle  2_{b}3_c4_d5_e\rangle[1_{\dot a}2_{\dot{b}}3_{\dot{c}}4_{\dot{d}}][5_{\dot e}|\sigma^\mu p_2 |1_{a}\rangle}{s_{34}}-\frac{\langle 1_a 2_{b}3_c4_d\rangle[2_{\dot{b}}3_{\dot{c}}4_{\dot{d}}5_{\dot{e}}][1_{\dot a}|p_2\sigma^\mu |5_{e}\rangle}{s_{34}}+(1\leftrightarrow 2 )\right]\\
\notag&&\hspace{-0.4cm}-\frac{\langle 2_b3_c4_d5_e\rangle[1_{\dot{a}}3_{\dot{c}}4_{\dot{d}}5_{\dot e}][2_{\dot{b}}|p_1\sigma^\mu p_5p_2|1_a\rangle}{s_{34}s_{12}}+\frac{\langle 1_a3_c4_d5_e\rangle[2_{\dot{b}}3_{\dot{c}}4_{\dot{d}}5_{\dot e}][1_{\dot{b}}|p_2 p_5\sigma^\mu p_1|2_b\rangle}{s_{34}s_{12}}
\,.
\eqae
Here we follow the convention for 6D spinor helicitiy formalism in \cite{Bern:2010qa}. The six-dimensional null-momentum can be parameterized by the six dimensional spinors
\eqa
P^{AB}=\lambda^{A a}\lambda^{B b}\epsilon_{ab}\,,~~P_{AB}=\tilde\lambda_{A \dot a}\tilde\lambda_{B\dot b}\epsilon^{\dot a\dot b}\,,
\eqae  
where $A,B$ are the SU*(4) Lorentz indices and $a$, $\dot a$ are the SU(2) little group indicies. We can express the six dimensional spinors in terms of massless four dimensional spinors as
\eq
\lambda^{A}_{~\,a}=\left(\begin{array}{cc}-\frac{ m}{\langle\lambda\mu\rangle}\mu_\alpha&\lambda_\alpha\\
\tilde\lambda^{\dot\alpha}&\frac{ \tilde m}{[\mu\lambda]}\tilde\mu^{\dot \alpha}\end{array}\right)
\,,~~~\tilde \lambda_{A\dot a}=\left(\begin{array}{cc}\frac{\tilde m}{\langle\lambda\mu\rangle}\mu^\alpha&\lambda^\alpha\\
-\tilde\lambda_{\dot\alpha}&\frac{  m}{[\mu\lambda]}\tilde\mu_{\dot \alpha}\,,
\end{array}\right),
\eqe 
where $\mu$, $\tilde \mu$ are the reference spinors.
To cast down to $(F_{1-}F_{2-}F_{3+}F_{4+}\phi_5)$, we choose $(a,b,c,d,e)=(1,1,2,2,1)$ and $(\dot{a},\dot{b},\dot{c},\dot{d},\dot{e})=(\dot{1},\dot{1},\dot{2},\dot{2},\dot{2})$. 
As a result, the five-point amplitude can be written as (with $m=\tilde{m}$)
\eq
\sum_{S_5/(Z_5\times Z_2)} I_{5} (1,2,3,4,5) (C\cdot\ell)
=
(m[12]^2\langle34\rangle^2) \sum_{S_5/(Z_5\times Z_2)} I_{5} (1,2,3,4,5)\,,
\eqe 
where we sum over all distinct permutations by mod out cyclic symmetry $Z_5$ and reflection symmetry $Z_2$, and $I_{5}(1,2,3,4,5)$ is the four-dimensional pentagon integral with massive propagators,
\eqa
I_{5} (1,2,3,4,5) &=& \int {d^4 \ell \over (2 \pi)^4} {1\over (\ell^2-m^2) ((\ell+k_1)^2 - m^2)
((\ell+k_1+k_2)^2 - m^2)} \cr
& \times & {1 \over ((\ell- k_5)^2 - m^2) ((\ell-k_4-k_5)^2 - m^2) } \, ,
\eqae
here the loop momentum $\ell$ is in four dimensions now. 

We integrate away the massive degrees of freedom by taking the large mass limit. In practice, this correspond to setting all $\ell\cdot k$ to zero in the propagators. In the leading order of large mass limit, 
\eq
I_{5} (1,2,3,4,5) \rightarrow \int {d^4 \ell \over (2 \pi)^4} {1\over (\ell^2-m^2)^5  }
= -i{1 \over 12(4 \pi)^2 m^6}\,,
\eqe
and thus we obtain the one-loop contribution to the operator 
\eq
 -i \frac{4!}{2}{1 \over 12 (4\pi)^2 m^6}\cdot (m[12]^2\langle34\rangle^2)=-i\frac{[12]^2\langle34\rangle^2}{ m^5(4\pi)^2}
 \,,
 \eqe
where the ${4!}/{2}$ arises from summing over permutations. We can compare this result with one-loop contribution to $F^4$ operator, which is given by a box integral with massive propagators,\footnote{The computations on $F^4$ and $D^4F^4$ were obtained together with Massimo Bianchi, Francisco Morales and Gaberiele Travaglini~\cite{Workinprogress}.}  
\eq
-s_{12}s_{14} A^{\rm tree}(1,2,3,4) \sum_{S_4/(Z_4\times Z_2)} I_{4} (1,2,3,4)\,,
\eqe
where the box integral $I_{4} (1,2,3,4)$ is defined as
\eq
I_{4} (1,2,3,4) = \int \frac{d^4\ell}{(2\pi)^4}\frac{1}{(\ell^2-m^2)((\ell+k_2)^2-m^2)((\ell+k_2+k_3)^2-m^2)((\ell-k_1)^2-m^2)} \, .
\eqe
Here $A^{\rm tree}(1,2,3,4)$ is the four-point tree-level gluon amplitude, choose the helicity configuration as that of $F^4\phi$ we discussed previously, we have
\eq
-s_{12}s_{14} A^{\rm tree}(1,2,3,4) = \langle 12 \rangle^2 [34]^2 \, .
\eqe
To the leading order, we obtain the one-loop contribution to the operator $F^4$
\eq
{3! \over 2 }[12]^2\langle34\rangle^2 \int \frac{d^4\ell}{(2\pi)^4}\frac{1}{[\ell^2-m^2]^4} = 
i \frac{[12]^2\langle34\rangle^2}{2 m^4(4\pi)^2} \, .
\eqe
Thus we find indeed the one-loop coefficients of $F^4\phi$ and $F^4$ is simply different by a factor of $-2$, and we know that the one-loop results are actually exact in all orders of coupling constant.  

We can also obtain the one-loop contributions to operators $D^{m} F^4$ and $D^{m} F^4 \phi$ by expanding the corresponding box and pentagon integrals to higher orders in $k_i$. It is easy to see any odd order in $k_i$ vanishes due to the integration $\ell$. The first non-trivial case is to the second order in $k_i$, which corresponds to operators $D^{2} F^4$ and $D^{2} F^4 \phi$. We find the results for both operators vanish on-shell, since they both are proportional to $\sum_{i<j} s_{ij}=0$. We then further expand to the fourth order in $k_i$, after the integration over $\ell$, we find for $D^{4} F^4$ the result is given by
\beq
i [12]^2\langle34\rangle^2 \left( \sum_{i<j} s^2_{ij} \right){1 \over  240 (4 \pi)^2  m^8 } \,,
\eeq
while for $D^{4} F^4 \phi$ it is
\eq
-i[12]^2\langle34\rangle^2 \left( \sum_{i<j} s^2_{ij} \right) {1 \over 120 (4\pi)^2 m^{10}} \, .
\eqe
Thus indeed the same as the case of $F^4$ and $F^4\phi$, we find the one-loop coefficients of $D^4F^4$ and $D^4F^4\phi$ differ by a factor of $-2$, although now generally they receive higher-loop as well as instanton corrections.

%%%%%%%%%%%%%%%%%%%%%%%%%%%%%%%%%%%%%%%%%%%%%%%%%%%%%%%%%%%%%%%%
\section{Three dimensions\label{3D}}
%%%%%%%%%%%%%%%%%%%%%%%%%%%%%%%%%%%%%%%%%%%%%%%%%%%%%%%%%%%%%%%%
In three dimensions, the little group is $Z_2$, and thus we only have bosons and fermions. Here we will again consider supersymmetry constraints on higher-dimensional operators, in particular six-derivative terms.  We will consider three cases, $\mathcal{N}=8$ with SO(8) and SO(7) R-symmetry, where the latter correspond to that of SYM, and the $\mathcal{N}=6$ SO(6) of Chern-Simons matter theories.  As we will demonstrate, the coefficient of the dimension $6$ operator $(\partial\phi)^6$ will again be completely determined by that of $(\partial\phi)^4$, for $\mathcal{N}>6$.  

Since now all bosonic states are equivalent to scalars, there is no corresponding helicity categorization. The on-shell representation of $\mathcal{N}={\rm even}$ supersymmetry is furnished by breaking SO($\mathcal{N}$) to U($\mathcal{N}/2$). Thus we introduce $\eta^I$ as fundamentals of U($\mathcal{N}/2$), and represent the SO($\mathcal{N}$)  generators as 
\eq\label{3DGen}
R^{IJ}=\eta^I\eta^J \, ,\quad R^I\,_J=\eta^I\partial_{\eta^J}-\frac{1}{2} \delta^I_J \, , \quad R_{IJ}=\partial_{\eta^I} \partial_{\eta^J}\,. 
\eqe
Due to the fact that the R-symmetry generators that are part of SO($\mathcal{N}$)/U($\mathcal{N}/2$) are non-linear, solutions to their constraints are rather involved. For $\mathcal{N}$=8 theories, the on-shell degrees of freedom are encoded in a single scalar superfield. In the case where only SO(7) is present, the R-symmetry generators are~\cite{LipsteinMason}:
\eq
R^{IJ}=\eta^I\eta^J+\frac{1}{2}\epsilon^{IJKL}\partial_{\eta^K}\partial_{\eta^L},\quad R^I\,_J=\eta^I\partial_{\eta^J}-\frac{1}{4}\delta^I_J \eta^K\partial_{\eta^K}\,,
\eqe
where now $I=1,\cdots,4$. For $\mathcal{N}=6$, it is contained in a scalar and a fermionic superfield, $\Phi$ and $\overline{\Psi}$ respectively, with their degrees of freedom contained as,
\eqa
\nonumber \Phi&=&\phi^4+\eta^I\psi_I+\eta^{I}\eta^J\phi_{IJ}+\eta^3 \psi_4 \, ,\\
\overline{\Psi}&=&\bar{\psi}^4+\eta^I\bar{\phi}_I+\eta^{I}\eta^J\bar{\psi}_{IJ}+\eta^3 \bar{\phi}_4\,.
\eqae
Note that due to the constant $\frac{3}{2}$ for the U(1) generator in eq.(\ref{3DGen}), for $\mathcal{N}=6$, the matrix element must be of even multiplicity.

We now consider the possible local dimension-six matrix elements. For maximal $\mathcal{N}=8$, we have:\footnote{Other invariants such as $(q^{A\alpha}_1q^B_{2\alpha}q^{C\alpha}_1q^D_{3\alpha} + {\rm perm})$ are related to the ansatz after permutation due to Schouten identities and super-momentum conservation.}
\eq\label{3DAnsatz1}
\delta^8(Q^{\alpha I})\epsilon_{ABCD}\left( q^{A\alpha}_1q^B_{2\alpha}q^{C\beta}_1q^D_{2\beta}+ {\rm perm} \right) \equiv \delta^8(Q)h(\lambda,\eta)\,,
\eqe  
This ansatz manifestly vanishes under the multiplicative SUSY generator as well as having the correct permutation invariance symmetry. It is required to be degree 12 in the $\eta$s, due to the U(1) generator $R^I_I$. It is rather non-trivial to check that it vanishes under the multiplicative R-symmetry generators. To simplify our task, we will follow~\cite{Till} and project the fermionic variables $\eta_i$ on a convenient basis.
  
We begin by introducing the following $n/2$-set of bosonic $n$-dimensional objects 
\eq
x^{\pm}_{\mathfrak{a}}, \quad y^{\alpha}
\eqe
where $\mathfrak{a}=1,2,\cdots,(n-4)/2$ and the new variables satisfies
\eq\label{XDef}
x_{\mathfrak{a}}^{\pm}\cdot\lambda=x_{\mathfrak{a}}^\pm \cdot x_{\mathfrak{b}}^\pm= x^\pm_{\mathfrak{a}} \cdot y^\alpha=0,\quad x_{\mathfrak{a}}^+\cdot x_{\mathfrak{b}}^-=\delta_{\mathfrak{a}\mathfrak{b}},\quad y^{\alpha}\cdot\lambda^{\beta}=\epsilon^{\alpha\beta}\,,
\eqe
The inner products in the above represent summing over all external leg labels, i.e. $\lambda\cdot\lambda=\sum_i\lambda_i\lambda_i$\,. At six points, an explicit solution to $x^\pm$ is given as:
\eqa
\nonumber x^{\pm}_{i}=\frac{\epsilon_{ijk}\langle jk\rangle}{2\sqrt{2}\sqrt{p^2_{123}}}, \quad i,j,k\in odd\\
x^{\pm}_{i}=\frac{\pm i\epsilon_{ijk}\langle jk\rangle}{2\sqrt{2}\sqrt{p^2_{123}}}, \quad i,j,k\in even\,.
\eqae
The explicit form of $y^\alpha$ is irrelevant as we will show that SUSY invariance dictate that the amplitude must be independent of it. 

The aim of introducing these variables is to separate the fermionic variables into pieces which vanish under the presence of $\delta^{\mathcal{N}}(Q^{\alpha I})=\delta^{\mathcal{N}}\left(\sum_i\lambda_i^{\alpha}\eta^I_{i}\right)$ . Defining:
\eq
\alpha_{\mathfrak{a}}^{\pm,I}=c^{\pm}_{\mathfrak{a}}\cdot \eta^I,\quad Y^{\alpha,I}=y^{\alpha }\cdot \eta^I\,,
\eqe
we see that 
\eq\label{Conversion}
\eta_{i}^I=\sum_{\mathfrak{a}=1}^{(n-4)/2}\left((x_i)_{\mathfrak{a}}^{+}\alpha_{\mathfrak{a}}^{-,I}+(x_i)_{\mathfrak{a}}^{-}\alpha_{\mathfrak{a}}^{+,I}\right)+\epsilon_{\alpha\beta}(\lambda_i^\alpha Y^{\beta,I}-y_i^{\alpha}Q^{\beta,I})
\eqe
and thus the last term would be the degree of freedom that is projected out by $\delta^{\mathcal{N}}(Q^{\alpha I})$. Now let's consider the constraint imposed by SUSY. The vanishing under the generator $Q^{\alpha I}$, is achieved in the usual way of requiring the amplitude to be of the form:
\eq
\mathcal{A}_n=\delta^{\mathcal{N}}(Q^{\alpha I})F(\lambda_i,\,\alpha_{a}^{\pm,I},\,Y^{\alpha,I})\,.
\eqe
The vanishing under $Q^{\alpha }_I=\lambda^\alpha\cdot \partial/\partial \eta^I$ the implies 
\eq
\lambda^\alpha\cdot \frac{\partial}{\partial \eta^I}F=\lambda^\alpha\cdot\left(x^+\frac{\partial}{\partial \alpha^{+,I}}+x^-\frac{\partial}{\partial \alpha^{-,I}}+y^{\alpha}\frac{\partial}{\partial Y^{\alpha,I}}\right)F=-\frac{\partial}{\partial Y^{\alpha,I}}F=0\,,
\eqe
thus one concludes that $F$ is independent of $Y^{\alpha,I}$\,. Finally, the non-linear R-symmetry generators now take the form:
\eq
R^{IJ}=\sum_{\mathfrak{a}}\alpha^{+[I}_{\mathfrak{a}}\alpha^{-J]}_{\mathfrak{a}},\quad R_{IJ}=\sum_{\mathfrak{a}}\frac{\partial}{\partial\alpha^{+[I}_{\mathfrak{a}}}\frac{\partial}{\partial\alpha^{-J]}_{a}}\,.
\eqe 
Thus constraints from SUSY and R-symmetry tells us that the six-point matrix element must take the form: 
\eq\label{Target}
\mathcal{A}_6=\delta^{8}(Q^{\alpha I})(\delta^{4}(\alpha^{+})f_+(\lambda)+\delta^{4}(\alpha^{-})f_-(\lambda))\,.
\eqe

Thus our task is to show that when re-expressing eq.(\ref{3DAnsatz1}) in terms of the new fermionic basis using eq.(\ref{Conversion}), it must take the form of eq.(\ref{Target}). However straight forward substitution would result in $h(\lambda, \eta)$ to contain terms proportional to $Y^m$ with $m<4$, which is forbidden under the full SO(8) symmetry. Thus one concludes that there are no dimension-six local matrix elements, and this implies that the $c_6$ must be proportional to $c^2_4$, where $c_n$ is the coefficient of the dimension $n$ operator. We finally comment that the procedure of introducing nice variables $\alpha^{\pm}$ should be very useful for higher-point analysis, for the case of six-point amplitude, one can actually simply test the validity of the SUSY generator $\lambda_{\alpha} \partial_{ \eta^A }$, and this is the approach we will take the amplitudes in 6D.

A similar analysis applies to $\mathcal{N}=8$ SYM with SO(7). Note that the SU(4) part of SO(7) allows the six-point matrix element to be of degree 8, 12, and 16 in $\eta$'s. For degree $8$, mass dimension 6 would require the local matrix element to have the form $\delta^{(8)}(Q)(s_{ij}+ {\rm perm} )$ which vanishes through momentum conservation. The lack of degree 8 matrix element also implies the same fate for degree 16, since the two are related via Fourier transform  in $\eta$. Finally the generator $R^{IJ}$ would then act solely on the degree 12 matrix element which can only vanish if  $\sum_i\eta_i^I\eta_i^J$ and $\sum_i\partial_{\eta_i^I}\partial_{\eta_i^J}$  vanishes separately, for which the former possibility is already ruled out from the previous analysis. Thus we can conclude that for $\mathcal{N}=8$ SYM, one the effective theory on the Coulomb branch again requires $c_6$ to be proportional to $c^2_4$.

It was argued in~\cite{DineSeiberg}, and computed explicitly in~\cite{3DInst1, 3DInst2, PSSF43D}, that the four derivative term of the effective action for $\mathcal{N}=8$ SYM receives one-loop and non-perturbative instanton corrections. Our result then implies that the dimension-six operator receives perturbative contribution only at two loops. While the one-instanton corrections come from the cross terms of one-loop and one-instanton contributions (and the perturbation around the one-instanton background) of dimension-four operator. More generally, the $n$-instanton contributions to the operator come from the cross terms with $k$ and $l$-instanton contributions of the dimension-four operator, with $k+l=n$ ($0$-instanton contribution means the perturbative one-loop contribution.).

%%%%%%%%%%%%%%%%%%%%%%%%%%%%%%%%%%%%%%%%%%%%%%%%%%%%%%%%%%%%%%%%
\section{Six dimensions\label{6D}}
%%%%%%%%%%%%%%%%%%%%%%%%%%%%%%%%%%%%%%%%%%%%%%%%%%%%%%%%%%%%%%%%
Here we consider six-dimensional theories with sixteen as well as eight supercharges. This includes the $\mathcal{N}=(2,0)$ and $\mathcal{N}=(1,0)$ theories of self-dual two forms, and the $\mathcal{N}=(1,1)$ SYM theory. The on-shell kinematic variables are~\cite{Donal}:
\eq
p^{AB}=\lambda^{Aa}\lambda^B\,_{a},\quad p_{AB}=\tilde{\lambda}_{\dot{a}A}\tilde{\lambda}^{\dot{a}}\,_B,\quad \lambda^{aA}\tilde{\lambda}_{\dot{a}A}=0\,,
\eqe
where $A,B$ are fundamental SO(1,5)$\sim$SU$^*$(4) Lorentz indices, whilst $(a,\dot{a})$ are SO(4)$\sim$ SU(2)$\times$SU(2) indices. The self-dual two-form and three-form field strengths are written as (explicit decomposition in terms of SU(4) irrep is give n in appendix~\ref{App1}):
\eq
F_{\mu\nu}\sim F^{A}\,_{B}=\lambda_a\,^{A}\tilde{\lambda}_{\dot{a},B},\quad H^+_{\mu\nu\rho}\sim H^{(AB)}=\lambda_{(a}\,^{A}\lambda_{b)}\,^{B}\,.
\eqe
Here, $H^+_{\mu\nu\rho}$ indicates it satisfies the self-duality condition $H_{\mu\nu\rho}=\frac{1}{3!}\epsilon_{\mu\nu\rho\sigma\tau\upsilon}H^{\sigma\tau\upsilon}$\,.

Supersymmetry is implemented by introducing the following Grassmann-odd variables~\cite{6DSUSY}: 
\eq
\mathcal{N}=(1,1),\quad (\eta_{a},\bar{\eta}_{\dot{a}}),\quad \mathcal{N}=(2,0),\quad (\eta^I_{a})\,.
\eqe
Here, $(\eta_{a},\tilde{\eta}_{\dot{a}})$ and $ (\eta^I_{a})$ with $I=1,2$, carries charge (+,0) and (0,+) under the Cartans of SU(2)$\times$SU(2) and Sp(4) R-symmetry respectively. As a result, the sixteen supercharges are given as:
\eqa
\nonumber \mathcal{N}=(1,1),&\quad& (Q^{+A},\,Q^{-A},\, \tilde{Q}^+_{A},\, \tilde{Q}^-_{A})=(\lambda^{aA}\eta_a, \,\lambda^{aA}\frac{\partial}{\partial\eta^a},\, \tilde\lambda_{\dot{a}A}\tilde{\eta}^{\dot{a}},\, \tilde\lambda_{\dot{a}A}\frac{\partial}{\partial\tilde{\eta}_{\dot{a}}})\,, \\
 \mathcal{N}=(2,0),&\quad& (Q^{+IA},\,Q^{-A}_I)=(\lambda^{aA}\eta^I_a, \,\lambda^{aA}\frac{\partial}{\partial\eta^{aI}})\,.
\eqae

Similar to three dimensions, the R-symmetry generators involve both linear and bi-linear operators in the $\eta$'s. The generators of SU(2)$\times $SU(2) for the $\mathcal{N}=(1,1)$ theory is given as:
\eq
\displaystyle{\left(\sum_i\eta_i^{a}\eta_{ia},\;\sum_{i}\eta_i^a\partial_{\eta_i^a},\; \sum_i\partial_{\eta_i^a}\partial_{\eta_{ia}}\right)\oplus\left( \eta \rightarrow \tilde\eta\right)}
\eqe
whilst the Sp(4) generators for $\mathcal{N}=(2,0)$ are given as:
\eq
\displaystyle{\left(\sum_i\eta_i^{Ia}\eta^{J}_{ia},\;\sum_{i}\eta_i^{Ia}\partial_{\eta_i^{Ja}},\; \sum_i\partial_{\eta_i^{Ia}}\partial_{\eta^J_{ia}}\right)}
\eqe
where the first and the last generator are that of Sp(2)$\times$Sp(2) subgroup.
%%%%%%%%%%%%%%%%%%%%%%%%%%%%%%%%%%%%%%%%%%%%%%%%%%%%%%%%%%%%%%%%
\subsection{SUSY completions of dimension-six operators}
%%%%%%%%%%%%%%%%%%%%%%%%%%%%%%%%%%%%%%%%%%%%%%%%%%%%%%%%%%%%%%%%
We now consider the possible supersymmetric completion of the on-shell matrix elements $H^6$ and $F^6$. The superfield expansion for the vector and the tensor multiplet is given as~\cite{Arthur}:
\begin{eqnarray}
\null\mathcal{N}=(1,1):\quad\Phi(\eta,\tilde{\eta}) &=&
  \phi 
  + \chi^a \eta_a 
  + \phi'(\eta)^2 
  + \tilde{\chi}_{\dot{a}}\tilde{\eta}^{\dot{a}} 
  + g^a\,_{\dot{a}}\eta_a\tilde{\eta}^{\dot{a}}
  + \tilde{\psi}_{\dot{a}}(\eta)^2\tilde{\eta}^{\dot{a}}\nonumber\\
  && \null
  + \phi''(\tilde{\eta})^2
  + \psi^a\eta_a(\tilde{\eta})^2
  + \phi'''(\eta)^2(\tilde{\eta})^2 \,,\nonumber\\\
  \mathcal{N}=(2,0):\quad\Phi(\eta^I) &=&
  \phi 
  + \chi_I^{a} \eta^I_a 
  + \phi_{IJ}\eta^{a(I}\eta^{J)}_a
  + b_{ab}\eta^{(aI}\eta^{b)}_I
  +\chi^J_{a} \eta^I_a\eta_{I}^b\eta^a_{J} 
  + \eta^4\bar{\phi}\,,\nonumber\\
  \end{eqnarray}
where $(\eta^2)^{(IJ)}\equiv\tfrac{1}{2}\epsilon^{ab}\eta^{I}_b\eta^{J}_a$. Since the two-form $b_{ab}$ and the vector $g^a\,_{\dot{a}}$ sits in the middle of the multiplet, an ansatz for dimension-six operators are given as:
\eqa \label{6Dansatz20}
\mathcal{N}=(2,0):\quad&&\delta^8(Q^{AI})\left(\,\langle q^I_1q_{2I}q^J_3q_{4J}\rangle + {\rm perm} \right)
\eqae
where $q^{A,I}_i=\lambda^{Aa}_i\eta^I_{ia}$ and $\langle q^I_1q_{2I}q^J_3q_{4J}\rangle=\epsilon_{ABCD}q^{AI}_1q^B_{2I}q^{CJ}_3q_{4J}^D$. First of all, note that since the latter is anti-symmetric under the exchange of any two $q_i$s, the ansatz for $\mathcal{N}=(2,0)$ is actually zero. Thus for $\mathcal{N}=(2,0)$, any local dimension-six operator must be cancelled by the factorization diagrams of two dimension-four operators, thus requiring $c_6\sim c^2_4$. This result was also found in~\cite{XYSG1}. 

For the $\mathcal{N}=(1,1)$ case, we have an ansatz that are parametrized by one degree of freedom,
\eq\label{6Dansatz11}
 \mathcal{N}=(1,1):\quad \delta^4(Q^{A})\delta^4(\tilde{Q}_{B})\left( \; q_1^A\tilde{q}_{2A}q_2^B\tilde{q}_{1B} + a_1q_1^A\tilde{q}_{2A}q_1^B\tilde{q}_{2B}+\mathrm{perm}\;\right)\,.
\eqe
Other seemingly different terms are actually related to these two listed above via super-momentum conservation. For instance $(q_1^A\tilde{q}_{2A}q_2^B\tilde{q}_{3B}+\mathrm{perm})$ can be recast into $(q_1^A\tilde{q}_{2A}q_2^B\tilde{q}_{1B}+\mathrm{perm})$, while $(q_1^A\tilde{q}_{2A}q_3^B\tilde{q}_{2B}+\mathrm{perm})$ can be expressed in terms of  $(q_1^A\tilde{q}_{2A}q_1^B\tilde{q}_{2B}+\mathrm{perm})$. Moreover, $(q_1^A\tilde{q}_{2A}q_3^B\tilde{q}_{4B}+\mathrm{perm})$ is a linear combination of $(q_1^A\tilde{q}_{2A}q_3^B\tilde{q}_{2B}+\mathrm{perm})$ and $(q_1^A\tilde{q}_{2A}q_1^B\tilde{q}_{2B}+\mathrm{perm})$. By applying the supersymmetry generator $\sum_i \lambda_i^A \cdot \frac{\partial}{\partial \eta_i}$, one can find the ansatz does not satisfy the supersymmetry constraint. Then we conclude there is no SUSY completion of local dimension-six operator for $\mathcal{N}=(1,1)$.\par

For the half-maximal SUSY, we again consider the tensor multiplet, where now the degrees of freedom are encoded in two fermionic superfields transforming as a doublet under the chiral SU(2) little group~\cite{Rozali}. It turns out that for $(1,0)$ supersymmetry there is only one possible independent ansatz that we can write down, 
\eq \label{6Dansatz10}
 \mathcal{N}=(1,0):\quad\delta^4(Q^{A}) \left(\, \langle q_1q_{2} 12\rangle\langle 3456\rangle + {\rm perm} \right)\,,
\eqe 
where now the permutation is summing over all antisymmetrization of external legs, this is due to the fermionic property of the superfield. Like the case of $(1,1)$ one may consider other possible terms, such as $\langle q_1 q_2 3 4\rangle \langle 1256\rangle + {\rm perm}$, $\langle q_1 q_2 1 3\rangle \langle 2456\rangle + {\rm perm}$ or $\langle q_1  1 3 4\rangle \langle q_2 2 56\rangle + {\rm perm}$, in fact they are not independent of the one we have considered in (\ref{6Dansatz10}). Again we check the validity of the operator by acting with the SUSY generator $\sum_{i} \lambda_i^{ A} \cdot \partial_{\eta_i}$, we find that the ansatz is not annihilated by the generator. Thus it is ruled out by the $(1,0)$ supersymmetry, which then requires that the the coefficient of the dimension-six operator is proportional to the square of that of the dimension-four operators.

%%%%%%%%%%%%%%%%%%%%%%%%%%%%%%%%%%%%%%%%%%%%%%%%%%%%%%%%%%%%%%%%
\section{Conclusions} \label{conclusion}
%%%%%%%%%%%%%%%%%%%%%%%%%%%%%%%%%%%%%%%%%%%%%%%%%%%%%%%%%%%%%%%%

In this paper, we apply the approach of scattering amplitudes to study how supersymmetry can constrain the effective actions of theories in various dimensions. In a simple but very efficient way, we derive non-renormalization theorems for four-dimensional theory with sixteen supercharges, it includes a particular class of abelian operators with respecting SU(4) R-symmetry, $(F_-)^2 (F_+)^{2m}$, or those operators break SU(4) to Sp(4), $(F_-)^2 (F_+)^{2m} \phi^m$. We find the coefficients of all these operators are completely determined by the corresponding lowest irrelevant operators, $(F_-)^2 (F_+)^{2}$ and $(F_-)^2 (F_+)^{2} \phi^m$. Using the known fact that operators $(F_-)^2 (F_+)^{2}$ and $(F_-)^2 (F_+)^{2} \phi$ are one-loop exact, and do not receive any non-perturbative instanton corrections, we thus conclude the same for the $(F_-)^2 (F_+)^{2q}$ and $(F_-)^2 (F_+)^{2q} \phi$ that they must be $q$-loop exact. 

We further extend our analysis to theories in other dimensions, including maximally supersymmetric theories in three dimensions, as well as theories in six dimensions with various choices of supersymmetries: $(2,0)$, $(1,0)$ and $(1,1)$. We find there are similar non-renormalization theorems, which relate the coefficients of dimension-six operators $H^6$ for $(2,0)$ and $(1,0)$ or $F^6$ for $(1,1)$ to those of dimension-four operators. 

The fact that we can give precise coefficients for an infinite set of operators in four dimensions, is largely due to the ability to organize our operators in terms of helicities, at the same time half of the SUSY generators are linearly realized. For higher dimensions in principle one can also choose to use representations organized in terms of a U(1) subgroup of the little group. However, in such representation, the SUSY charges are given as linear combinations of derivative and multiplicative generators. This results in SUSY invariants that contain different degree of Grassmann polynomials, which are difficult to analyze. One can instead use R-symmetry decomposition instead much like the $x^\pm$ variables introduced in three-dimensions~\cite{Till}. Explicit solutions for these projection variables at higher points may reveal new protected sectors.  

It would be very interesting to extend our analysis for the SYM effective action in a non-abelian background $F_{mn}$. In the non-abelian case, now one needs to consider non-vanishing commutator $[F,F]$, and take in account the fact that $[D, D] \sim F$. Since scattering amplitudes are free of the redefinition of fields, so our approach should have the advantage to study the effective actions with non-abelian gauge groups as well. 

%%%%%%%%%%%%%%%%%%%%%%%%%%%%%%%%%%%%%%%%%%%%%%%%%%%%%%%%%%%%%%%%
\section{Acknowledgement}
%%%%%%%%%%%%%%%%%%%%%%%%%%%%%%%%%%%%%%%%%%%%%%%%%%%%%%%%%%%%%%%%  

We would like to thank Heng-Yu Chen, Francesco Fucito, Juan Maldacena, Jan Troost and Xi Yin for useful communications, C.W would also like to thank Massimo Bianchi, Francisco Morales and Gaberiele Travaglini for helpful discussions and collaborations on related topics. Y-t. Huang and W-M. Chen is supported by MOST under the grant No. 103-2112-M-002-025-MY3. W-M. Chen is also supported by MOST under the grant No. 104-2917-I-002-014.

\begin{appendices}
%%%%%%%%%%%%%%%%%%%%%%%%%%%%%%%%%%%%%%
\section{Convection}\label{App1}
%%%%%%%%%%%%%%%%%%%%%%%%%%%%%%%%%%%%%%%
Our convention for spinor helicity formalism in four and six dimensions is summarised in the following. 
\begin{itemize}
\item Four dimensions
\end{itemize}
The definition of $\sigma$-matrices are
\eq
(\sigma^\mu)_{a\dot a}=(1,\vec{\sigma})\,,~~~(\bar\sigma^\mu)^{a\dot a}=(1,-\vec{\sigma})\,,
\eqe
where
\eq
\label{SGM}\sigma^0=\left(\begin{array}{rr}
1&\,\,~0\\
0&1
\end{array}\right)\,,~~
\sigma^1=\left(\begin{array}{rr}
0&\,\,~1\\
1&0
\end{array}\right)\,,~~\sigma^2=\left(\begin{array}{rr}
0&\,-i\\
i&0
\end{array}\right)\,,~~\sigma^3=\left(\begin{array}{rr}
1&0\\
0&-1
\end{array}\right)\,.
\eqe
The $\sigma$-matrices follow the completeness relation, with the metric $\eta=$diag$(+,-,-,-)$, 
\eq
\mathrm{Tr}\left[\left(\frac{\sigma^\mu}{\sqrt{2}}\right)\left(\frac{\bar \sigma^\nu}{\sqrt{2}}\right)\right]=\eta^{\mu\nu}\,,~~\left(\frac{\sigma^\mu}{\sqrt{2}}\right)_{a\dot a}\left(\frac{\bar \sigma_\mu}{\sqrt{2}}\right)^{\dot b b}=\delta_a^b \delta_{\dot a}^{\dot b}\,,
\eqe
which can be used to convert Lorentz four-vector indices to bi-spinor indices. As a result, one can show
\eq
\epsilon_{a\dot a,b\dot b,c\dot c,d\dot d}\equiv\epsilon^{\mu\nu\rho\lambda}\left(\frac{ \sigma_\mu}{\sqrt{2}}\right)_{a\dot a}\left(\frac{ \sigma_\nu}{\sqrt{2}}\right)_{b\dot b}\left(\frac{ \sigma_\rho}{\sqrt{2}}\right)_{c\dot c}\left(\frac{ \sigma_\lambda}{\sqrt{2}}\right)_{d\dot d}=i(\epsilon_{ab}\epsilon_{cd}\epsilon_{\dot d\dot a}\epsilon_{\dot b\dot c}-\epsilon_{da}\epsilon_{bc}\epsilon_{\dot a\dot b}\epsilon_{\dot c\dot d})\,,
\eqe
where $\epsilon^{0123}=1$, $\epsilon^{12}=\epsilon^{\dot 1\dot 2}=-\epsilon_{12}=-\epsilon_{\dot 1\dot 2}=1$.
Self- and anti-self-duality conditions for a tensor $f$ can be expressed in spinor indices,
\eq
f_{a\dot a,b\dot b}=\pm\frac{i}{2}\epsilon_{a\dot a,b\dot b,c\dot c,d\dot d}f^{c\dot c,d\dot d}\,,
\eqe
where the solutions to the equation with the plus and minus signs correspond to self-dual and anti-self-dual tensors, respectively. In general, 
a self-dual tensor can be written as
\eq
f_{a\dot a,b\dot b}=\epsilon_{ab}s_{\dot a\dot b},
\eqe
where the two indices in $s$ is symmetric.  The anti-self-dual tensor can be obtained by interchanging the dotted and undotted indices. 
\begin{itemize}
\item Six dimensions
\end{itemize}
Here we follows the convention in \cite{Donal}. The matrices $\Sigma$, $\tilde \Sigma$ satisfy Clifford algebra 
\eq
\Sigma^\mu \tilde \Sigma^\nu+\Sigma^\nu \tilde \Sigma^\mu=2\eta^{\mu\nu}
\eqe
and form a set of completeness relation as in four dimensions
\eq
\begin{array}{rclrcl}
\Sigma^\mu_{AB}\Sigma_{\mu CD}&=&-2\epsilon_{ABCD}\,,&~~~\Sigma^\mu_{AB}\tilde\Sigma_{\mu}^{ CD}&=&-2(\delta^C_A\delta^D_B-\delta^D_A\delta^C_B)\,,\\
&&&&&\\
\tilde\Sigma^{\mu AB}\tilde\Sigma_{\mu}^{ CD}&=&-2\epsilon^{ABCD}
\,,&~~~\mathrm{tr}\Sigma^\mu\tilde\Sigma^\nu&=&4\eta^{\mu\nu}\,,
\end{array}
\eqe
where $\mu=0,1,\dots,6$ are the SO(6) Lorentz indices and $A,B,C,D=1,2,3,4$ are the SU(4) spinor indices. 
One can use the completeness relation to obtain Levi-Civita tensor in spinor indices 
\eqa
\notag&&\hspace{-1cm}\epsilon_{A_1A_2,B_1B_2,C_1C_2,D_1D_2,E_1E_2,F_1F_2}\\
\notag&&=
-\epsilon_{A_1 E_2 F_1 F_2} \epsilon_{A_2 B_1 B_2 C_2} \epsilon_{C_1 D_1 D_2 E_1}+\epsilon_{A_1 B_1 B_2 C_2} \epsilon_{A_2 E_2 F_1 F_2} \epsilon_{C_1 D_1 D_2 E_1}\\
\notag&&~~~+\epsilon_{A_1 E_1 F_1 F_2} \epsilon_{A_2 B_1 B_2 C_2} \epsilon_{C_1 D_1 D_2 E_2}-\epsilon_{A_1 B_1 B_2 C_2} \epsilon_{A_2 E_1 F_1 F_2} \epsilon_{C_1 D_1 D_2 E_2}\\
\notag&&~~~+\epsilon_{A_1 E_2 F_1 F_2} \epsilon_{A_2 B_1 B_2 C_1} \epsilon_{C_2 D_1 D_2 E_1}-\epsilon_{A_1 B_1 B_2 C_1} \epsilon_{A_2 E_2 F_1 F_2} \epsilon_{C_2 D_1 D_2 E_1}\\
\notag&&~~~-\epsilon_{A_1 E_1 F_1 F_2} \epsilon_{A_2 B_1 B_2 C_1} \epsilon_{C_2 D_1 D_2 E_2}+\epsilon_{A_1 B_1 B_2 C_1} \epsilon_{A_2 E_1 F_1 F_2} \epsilon_{C_2 D_1 D_2 E_2}\\
\notag&&~~~-\epsilon_{A_1 A_2 B_2 F_2} \epsilon_{B_1 C_1 C_2 D_2} \epsilon_{D_1 E_1 E_2 F_1}+\epsilon_{A_1 A_2 B_1 F_2} \epsilon_{B_2 C_1 C_2 D_2} \epsilon_{D_1 E_1 E_2 F_1}\\
\notag&&~~~+\epsilon_{A_1 A_2 B_2 F_1} \epsilon_{B_1 C_1 C_2 D_2} \epsilon_{D_1 E_1 E_2 F_2}-\epsilon_{A_1 A_2 B_1 F_1} \epsilon_{B_2 C_1 C_2 D_2} \epsilon_{D_1 E_1 E_2 F_2}\\
\notag&&~~~+\epsilon_{A_1 A_2 B_2 F_2} \epsilon_{B_1 C_1 C_2 D_1} \epsilon_{D_2 E_1 E_2 F_1}-\epsilon_{A_1 A_2 B_1 F_2} \epsilon_{B_2 C_1 C_2 D_1} \epsilon_{D_2 E_1 E_2 F_1}\\
&&~~~-\epsilon_{A_1 A_2 B_2 F_1} \epsilon_{B_1 C_1 C_2 D_1} \epsilon_{D_2 E_1 E_2 F_2}+\epsilon_{A_1 A_2 B_1 F_1} \epsilon_{B_2 C_1 C_2 D_1} \epsilon_{D_2 E_1 E_2 F_2}\,.
\eqae
Then the (anti-)self-duality condition for a tensor $H$ can be written in spinor indices  
\eq
H_{A_1A_2,B_1B_2,C_1C_2}=\pm \frac{1}{2^3 3!}\epsilon_{A_1A_2,B_1B_2,C_1C_2,D_1D_2,E_1E_2,F_1F_2}H^{D_1D_2,E_1E_2,F_1F_2}\,,
\eqe
where the solution to the equation with plus and minus sign correspond to self-dual and anti-self-dual tensors. The general form of self-dual field strength is
\eq
H^{A_1A_2,B_1B_2,C_1C_2}_{sf}=h^{A_2B_2} \epsilon^{A_1B_1C_1C_2}-h^{A_2B_1} \epsilon^{A_1B_2C_1C_2}-h^{A_1B_2} \epsilon^{A_2B_1C_1C_2}+h^{A_1B_1} \epsilon^{A_2B_2C_1C_2}\,,
\eqe
and that of anti-self-dual field strength is
\eq
H_{asf}^{A_1A_2,B_1B_2,C_1C_2}=h_{GH} (\epsilon^{A_1A_2C_1G} \epsilon^{B_1B_2C_2H}-\epsilon^{A_1A_2C_2G} \epsilon^{B_1B_2C_1H})\,,
\eqe
where $h_{AB}$ and $h^{AB}$ are symmetric tensors in $A$ and $B$. Although $H_{sf}$ and $H_{asf}$ is not manifestly antisymmetric in the pairs of indices $\{A_i\}$, $\{B_i\}$, $\{C_i\}$, one can show the explicitly antisymmetric property by the identities 
\eq
\epsilon_{ABCD}\delta_{E}^F+\epsilon_{BCDE}\delta_{A}^F+\epsilon_{CDEA}\delta_{B}^F+\epsilon_{DEAB}\delta_{C}^F+\epsilon_{EABC}\delta_{D}^F=0\,,
\eqe
and
\eq
\epsilon_{ABCD}\epsilon^{EFGH}=4!\delta^E_{[A}\delta^F_B\delta^G_C\delta^H_{D]}\,,
\eqe
where the notation of antisymmetrization in indicies is defined as 
\eq
T_{[A_1\dots A_n]}=\frac{1}{n!}\sum_\sigma\mathrm{sgn}(\sigma)T_{A_{\sigma_1}A_{\sigma_2}\dots A_{\sigma_n}}\,,
\eqe
the sum here is for all permutations of indicies with the signature of permutation $\sigma$.

%%%%%%%%%%%%%%%%%%%%%%%%%%%%%%%%%%%%%%
\section{Solution to NMHV six-point SUSY Ward identity}\label{SUSYWard}
%%%%%%%%%%%%%%%%%%%%%%%%%%%%%%%%%%%%%%%

In this appendix we apply supersymmetric Ward identity to show that one can express full NMHV six-point superamplitude in terms of pure gluon amplitudes. First of all, as we discussed the four-point superamplitude is completely determined by little group scaling and its mass dimension, which is given by
\eq
\mathcal{A}_{4} = \delta^8(Q) { [34]^2 \over \langle 12\rangle^2 } \,,
\eqe
where the supercharge $Q^{\alpha, A}  = \sum^n_{i=1} \lambda^{\alpha}_i \eta^A_i$. It is straightforward to check that it produces the correct amplitude for $(F_-)^2(F_+)^2$. By projecting to the pure fermionic component, we obtain
\eq
A_{4}( \psi^{1}_1, \bar{\psi}^{234}_2, \psi^{1}_3, \bar{\psi}^{234}_4) = \langle 13 \rangle [24] s_{13} \, ,
\eqe 
where we have specified the SU(4) R-symmetry indices. We find the superamplitude is coincide with the four-point amplitude of Volkov-Akulov theory computed in~\cite{Chen:2014xoa}. Indeed DBI action with $\mathcal{N}=4$ supersymmetry completion contains Volkov-Akulov action. 

Let us now consider the six-point NMHV amplitude, from supersymmetric Ward identity~\cite{ElvangSUSYW}, the superamplitude can be expressed in terms of a basis with the coefficients being component amplitudes, 
 \begin{eqnarray}
\mathcal{A}_{6} &=& 
A(-+++--)X_{1111}
+A(\psi^{123} \psi^4 + + --)X_{1112} 
\cr
&+& {1 \over 2} A(s^{12} s^{34} + + --)X_{1122} 
+ ( 1 \leftrightarrow 2 )\, ,
\end{eqnarray}
where the basis $X_{ijkl}$ is defined as
\begin{equation}
X_{ijkl} = { m_{i n-3 n-2; 1} m_{j n-3 n-2; 2} m_{k n-3 n-2; 3} m_{l n-3 n-2; 4}  \over [n{-}3 \, n{-}2]^4 \langle n{-}1 n\rangle^4 } \delta^8(Q)
\end{equation}
with indices $i,j,k$ and $l$ symmetrized, and $m$ is defined as 
\begin{equation}
m_{i jk; A} = [ij] \eta^A_k + [jk] \eta^A_l + [kl] \eta^A_j
\end{equation}
Write it out explicitly for six points, 
\begin{eqnarray}  \label{SWI}
\mathcal{A}_{6} &=& 
A(-+++--){ \delta^4( [13]\eta_4 + \cdots ) \over [34]^4 \langle 56\rangle^4 } 
+A(\psi^{123} \psi^4 + + --) { \delta^3( [13]\eta_4 + \cdots )\delta( [23]\eta_4 + \cdots ) \over [34]^4 \langle 56\rangle^4 } 
\cr
&+& {1 \over 2} A(s^{12} s^{34} + + --){ \delta^2( [13]\eta_4 + \cdots )\delta^2( [23]\eta_4 + \cdots ) \over [34]^4 \langle 56\rangle^4 }  
+ ( 1 \leftrightarrow 2 )\, ,
\end{eqnarray}
where ``$\cdots$" denotes summing over cyclic permutations. From the above supersymmetric Ward identity (SWI), we can project it to pure gluon amplitudes, and obtain following linear equations 
\begin{eqnarray}
A(+++---) &=& 
{ [13]^4 \over [34]^4 } A(-+++--) 
+ { [13]^3[23] \over [34]^4  }  A(\psi^{123} \psi^4 + + --) \cr
&+& { [13]^2[23]^2 \over 2[34]^4  } A(s^{12} s^{34} + + --)  \cr
&+& ( 1 \leftrightarrow 2 ) \cr
A(++-+--) &=& 
{ [14]^4 \over [34]^4 } A(-+++--) 
+ { [14]^3[24] \over [34]^4  }  A(\psi^{123} \psi^4 + + --)  \cr
&+& { [14]^2[24]^2 \over 2[34]^4  } A(s^{12} s^{34} + + --)  \cr
&+& ( 1 \leftrightarrow 2 )\cr
A(--+++-) &=& 
{ \langle 26\rangle^4 \over \langle 56\rangle^4 } A(-+++--) 
- { \langle 26\rangle^3\langle 16\rangle \over \langle 56\rangle^4 }  A(\psi^{123} \psi^4 + + --) \cr
&+& { \langle 16\rangle^2 \langle 26\rangle^2 \over 2\langle 56\rangle^4 } A(s^{12} s^{34} + + --)  \cr
&+& ( 1 \leftrightarrow 2 )\, .
\end{eqnarray}
Since these three equations are independent, thus SWI allows one to solve $A(\psi^{123} \psi^4 + + --), A(s^{12} s^{34} + + --)$ and $A(\psi^{1} \psi^{234} + + --)$ in terms gluon amplitudes! Thus since for $F^6$ the NMHV amplitude $A(+++---)$, we find all other component amplitudes vanish as well. To be complete, the solution of above linear equations is given by
\begin{eqnarray}
\label{solutions}
A(\psi^{123} \psi^4 + + --) &=&
{[2 4] [3 4]^3 \langle 1 6 \rangle \over 
[1 2] [1 3] [2 3] \langle 6|4+5| 3] }A_1 + { [2 3] [3 4]^3 \langle 1 6 \rangle \over [1 4] [2 4] [1 2] \langle 6| 1+2 |4] } A_2 \cr
&+& { [2 3] [2 4] \langle 5 6 \rangle^4 \over 
\langle 6 1 \rangle \langle 2 6 \rangle \langle 6| 1+2 |3]  \langle 6| 1+2 |4]} A_3 \cr
A(s^{12} s^{34} + + --) &=&
{ [3 4]^3 ([2 4] \langle 2 6 \rangle -[1 4] \langle 1 6 \rangle  ) \over 
[1 2] [1 3] [2 3]  \langle 6|1+2| 3] } A_1 + {[3 4]^3 ([2 3] \langle 2 6 \rangle -[1 3] \langle 1 6 \rangle ) \over 
[1 2] [1 4] [2 4] \langle 6| 1+2| 4]} A_2 \cr
&+& { ([1 4] [2 3] + [1 3] [2 4]) \langle 5 6 \rangle^4 \over 
\langle 16 \rangle \langle 2 6 \rangle \langle 6| 1+2 |3]  \langle 6| 1+2 |4]} A_3 \cr
A(\psi^{1} \psi^{234} + + --) &=&
{ [1 4] [3 4]^3 \langle 2 6 \rangle \over 
[12][1 3] [2 3]  \langle 6| 1+2 |3] } A_1 + { [1 3] [3 4]^3 \langle 6 2 \rangle \over [1 2] [1 4] [2 4]  \langle 6| 1+2 |4] } A_2 \cr
&+& { [1 3] [1 4] \langle 5 6 \rangle^4 \over 
\langle 6 1 \rangle \langle 2 6 \rangle \langle 6| 1+2 |3]  \langle 6| 1+2 |4]} A_3
\end{eqnarray}
where $A_1, A_2$ and $A_3$ are defined as
\begin{eqnarray}
A_1 &=& A(+++---) - {[13]^4 \over [34]^4 } A(-+++--) - {[23]^4 \over [34]^4 } A(+-++--) \, , \cr
A_2 &=& A(++-+--) - {[14]^4 \over [34]^4 } A(-+++--) - {[24]^4 \over [34]^4 } A(+-++--) \, ,\cr
A_3 &=& A(--+++-) - { \langle 26\rangle^4 \over \langle 56\rangle^4 } A(-+++--)- { \langle 16\rangle^4 \over \langle 56\rangle^4 } A(+-++--) \, .
\end{eqnarray}
we have checked that all the unphysical poles amplitudes solved in (\ref{solutions}) cancel out. From this result, we can also obtain six-point NMHV amplitude for $\mathcal{N}=4$ SUSY completion of DBI action, whose gluon amplitude $A(+++---)$ is known~\cite{Rutger}, which can be computed, for instance, using CSW rules~\cite{Cachazo:2004kj},
\begin{eqnarray}
A(+++---) = { \langle 56\rangle^2 [12]^2 \langle 4 | 5+6 |3]^2 \over s_{124}} + {\rm permutations }\,,
\end{eqnarray}
where the permutations are summing over $1,2,3$ and $4,5,6$. We have checked that all the unphysical poles in the amplitudes solved in~(\ref{solutions}) cancel out. Now, we plug the solutions back to obtain the superamplitude in terms of pure gluon amplitudes only. We have checked numerically the superamplitude we obtain produce all correct component amplitudes, in particular it reproduces the single- and double-soft limits of DBI action~\cite{Cachazo:2015ksa} as well as that of the Volkov-Akulov theory~\cite{Chen:2014xoa}. 
\end{appendices}
%%%%%%%%%%%%%%%%%%%%%%%%%%%%%%%%%%%%%%%%%%%%%%%%%%%%%%%%%%%%%%%%
%%%%%%%%%%%%%%%%%%%%%%%%%%%%%%%%%%%%%%%%%%%%%%%%%%%%%%%%%%%%%%%%

\end{large}

\end{document}